\documentclass[12pt,preprint]{aastex}
\def\degpoint{\ifmmode ^{\rm{o}}\!. \else $^{\rm{o}}\!.$\fi}

\newcommand{\ms}{\mbox{m\,s$^{-1}$}}

\newcommand{\Msun}{\mbox{M$_{\odot}$}}

\newcommand{\Mjup}{\mbox{M$_{\rm Jup}$}}

\newcommand{\ltsimeq}{\raisebox{-0.6ex}{$\,\stackrel
         {\raisebox{-.2ex}{$\textstyle <$}}{\sim}\,$}}
\newcommand{\gtsimeq}{\raisebox{-0.6ex}{$\,\stackrel
         {\raisebox{-.2ex}{$\textstyle >$}}{\sim}\,$}}

\begin{document}

\title{Forever alone? Testing single eccentric planetary systems for 
multiple companions }

\author{Robert A.~Wittenmyer\altaffilmark{1}, Songhu 
Wang\altaffilmark{2}, Jonathan Horner\altaffilmark{1}, 
C.G.~Tinney\altaffilmark{1}, R.P.~Butler\altaffilmark{3}, 
H.R.A.~Jones\altaffilmark{4}, S.J.~O'Toole\altaffilmark{5}, 
J.~Bailey\altaffilmark{1}, B.D.~Carter\altaffilmark{6}, 
G.S.~Salter\altaffilmark{1}, D.~Wright\altaffilmark{1}, Ji-Lin 
Zhou\altaffilmark{2} }

\altaffiltext{1}{Department of Astrophysics, School of Physics, Faculty 
of Science, The University of New South Wales, Sydney, NSW 2052, Australia}
\altaffiltext{2}{Department of Astronomy \& Key Laboratory of Modern 
Astronomy and Astrophysics in Ministry of Education, Nanjing University, 
Nanjing 210093, China.}
\altaffiltext{3}{Department of Terrestrial Magnetism, Carnegie
Institution of Washington, 5241 Broad Branch Road, NW, Washington, DC
20015-1305, USA}
\altaffiltext{4}{University of Hertfordshire, Centre for Astrophysics
Research, Science and Technology Research Institute, College Lane, AL10
9AB, Hatfield, UK}
\altaffiltext{5}{Australian Astronomical Observatory, PO Box 915,
North Ryde, NSW 1670, Australia}
\altaffiltext{6}{Faculty of Sciences, University of Southern Queensland,
Toowoomba, Queensland 4350, Australia}

\email{
rob@phys.unsw.edu.au}

\shortauthors{Wittenmyer et al.}

\begin{abstract}

\noindent Determining the orbital eccentricity of an extrasolar planet 
is critically important for understanding the system's dynamical 
environment and history.  However, eccentricity is often poorly 
determined or entirely mischaracterized due to poor observational 
sampling, low signal-to-noise, and/or degeneracies with other planetary 
signals.  Some systems previously thought to contain a single, 
moderate-eccentricity planet have been shown, after further monitoring, 
to host two planets on nearly-circular orbits.  We investigate published 
apparent single-planet systems to see if the available data can be 
better fit by two lower-eccentricity planets.  We identify nine 
promising candidate systems and perform detailed dynamical tests to 
confirm the stability of the potential new multiple-planet systems.  
Finally, we compare the expected orbits of the single- and double-planet 
scenarios to better inform future observations of these interesting 
systems.

\end{abstract}


\section{Introduction}

The radial-velocity method remains the most versatile technique for 
determining the full orbital properties (e.g.~minimum mass, 
eccentricity) of extrasolar planets.  More than 500 planets have been 
discovered by this method, and considerable efforts are afoot to 
understand the true underlying distribution of planetary properties 
based on these data \citep{cumming08, shen08, otoole09a, monster, 
howard10, foreverpaper, jupiters, etaearth}.  However, there are 
significant biases in the measurement of planetary parameters from 
sparsely sampled radial-velocity data, especially if the amplitude ($K$) 
of the signal is small.  Particularly problematic is the orbital 
eccentricity, a quantity which is critically important for understanding 
the formation and dynamical history of planetary systems \citep{zhou07, 
ford08, shen08}.  Keplerian orbit-fitting algorithms, in a blind 
mathematical attempt to minimise $\chi^2$, often resort to increasing 
the eccentricity of a model fit.  \citet{shen08} found this introduced 
bias toward higher fitted eccentricities to be most egregious for data 
with signal-to-noise ratio $K/\sigma\,\ltsimeq$3.  Similarly, 
\citet{otoole09a} showed that there is a computational bias against 
fitting eccentricities near zero, and that the true uncertainties in 
orbital parameters can be 5-10 times larger than the formal 
uncertainties emerging from standard least-squares fits.  There is, 
therefore, reason to suspect that the observed eccentricity distribution 
of radial-velocity detected exoplanets is biased towards higher 
eccentricities.

Recently, further radial-velocity monitoring has revealed additional 
planets in two systems previously thought to host a single planet 
(HD\,142 and HD\,159868: Wittenmyer et al.~2012b), and the best-fit 
eccentricities of the previously known planets have significantly 
decreased as a result of fitting \textit{two} planetary signals.  This 
was most obvious for HD\,159868, where the previously known planet, with 
a period of $\sim$1180 days, had an orbit best fit with a very high 
$e=0.69\pm0.02$ \citep{otoole07}.  As a result of new data, and a 
two-planet solution, HD\,159868b is now best fit with a circular orbit 
($e=0.01\pm$0.03).  The possibility that two nearly-circular orbits can 
masquerade as a single, eccentric orbit has been explored by 
\citet{ang10} and \citet{rh09}.  Motivated by these findings, we now ask 
``Which eccentric single-planet systems can be better fit with two 
low-eccentricity planets?''

The ambiguities in orbital eccentricity can arise from one of four 
degeneracies, which we summarise here.  First, there is the degeneracy 
betweeen a single planet on an eccentric orbit and two planets on 
circular orbits in a 2:1 configuration \citep{ang10}.  Second is the 
degeneracy between one eccentric planet and two co-orbital planets (i.e. 
in a 1:1 resonance or ``Trojan pair''), as described in 
\citet{laughlin02} and \citet{g12}.  Third is the degeneracy between a 
single eccentric planet and a circular planet with a long-period 
companion \citep{rh09}.  The fourth degeneracy is that noted above for 
HD\,159868: that between a single eccentric planet and two 
nearly-circular planets with poorly sampled orbital periods 
\citep{142paper}.

This paper is organised as follows: Section 2 describes the data and 
analysis procedures, Section 3 gives the results of our efforts to fit 
two low-eccentricity planets, and details those systems which had the 
most promising results.  Finally, we discuss our conclusions in Section 
4.

\section{Data Analysis and Orbit Fitting}

We selected all radial-velocity detected single-planet systems with 
publicly available data and published eccentricities $e>0.3$.  This can 
be considered a high-eccentricity subset, as the mean eccentricity for 
the current population of confirmed planets is 0.22\footnote{Planet data 
obtained from the Exoplanet Orbit Database at http://exoplanets.org}.  
The mean uncertainty on the published eccentricities is 0.05, though we 
note that the uncertainties arising from least-squares fit can be 
underestimated \citep{otoole09a}.  In addition, we excluded any 
transiting planets (e.g.~HD\,17156b and HD\,80606b), because for these 
cases, the transit should be simultaneously fit with the radial 
velocities, a task which is beyond the scope of this paper.  After 
applying these selection criteria, 82 stars remained.  A summary of the 
data used here is given in Table~\ref{datasummary}.  All previously 
unpublished AAT data used in this work are now given in the Appendix, 
Tables~\ref{2039vels}--\ref{216437vels}.


To facilitate the comparison of the two-planet models with the single 
eccentric-planet model, we first re-fit all available radial-velocity 
data with a single planet (with no restrictions on $e$).  For those 
stars with data from multiple sources, this approach ensures consistent 
treatment by using the same fitting procedure for all stars.  These 
results are referred to as Method 1 as given in Table~\ref{allfits}.  
Then, for each system, we fit all available data using a genetic 
algorithm, which has proven useful in previous work where the system 
parameters are extremely uncertain or data are sparse \citep{NNSer, 
HUAqr, 47205paper, tinney11, cochran07}.  The genetic algorithm used 
here has the advantage that the range of allowed parameter space can be 
restricted: in this work, we wish to fit the data with two 
\textit{low-eccentricity} planets.  We thus direct the fitting process 
to model two Keplerian orbits with $e<0.2$.  We note in passing that 
applying this procedure to the data published in \citet{otoole07} yields 
a 2-planet system essentially identical to that presented in 
\citet{142paper}.  The best-fit set of parameters\footnote{While genetic 
algorithms are commonly lauded as ways to find a \textit{global} best 
fit, we note that their effectiveness depends on the choice of input 
parameters such as mutation rates and the degree to which $\chi^2$ is 
allowed to increase between generations. Hence, the ``global'' solutions 
found by our approach cannot necessarily be guaranteed to be the 
absolute best-fit for the the very complicated, multi-modal parameter 
space of the second planet.} resulting from 10,000 iterations of the 
genetic algorithm is then used as initial input for the 
\textit{Systemic} Console \citep{mes09}.  We then use \textit{Systemic} 
to perform Keplerian model fits to the data, again requiring $e<0.2$ for 
both planets.  These results are referred to as Method 2 as given in 
Table~\ref{allfits}.

The 12 systems for which a 2-planet fit gave a physically meaningful 
result (i.e.~no crossing orbits) with $\chi^2$ similar to the one-planet 
fit were subjected to a more detailed fitting process.  We used the 
Runge-Kutta algorithm within \textit{Systemic} to perform a dynamical 
fit which accounted for the gravitational interactions between the 
modelled planets.  The model systems were then integrated for $10^5$ 
years as a basic stability check.  The results of this analysis, those 
nine systems which remained stable for $10^5$ yr, are given in 
Table~\ref{promising}.  For those systems which proved stable in the 
initial check, we produced detailed dynamical maps of a broad range of 
parameter space about the best-fit orbits.  For this final step in the 
dynamical feasibility testing, we turned to the \textit{Mercury} N-body 
integrator \citep{chambers99}.  Following our previous work in dynamical 
mapping of extrasolar planetary systems \citep{24Sex, NNSer, HWVir, 
155358, 204313, 142paper, HUAqr}, we sampled a 3$\sigma$ region of 
four-dimensional parameter space: semimajor axis ($a$), eccentricity 
($e$), mean anomaly ($M$), and argument of periastron ($\omega$).  Due 
to computational limitations and the large uncertainties involved for 
these speculative 2-planet systems, we chose coarse grids: 21x21x5x5 in 
($a, e, M, \omega$), respectively.  As in previous work, we held the 
best-constrained planet (planet 1 as given in Table~\ref{promising}) 
fixed and altered the initial parameters of the second planet.  We ran 
each simulation for $10^8$ yr, or until the system destabilised (via 
ejection or collision).

\section{Results}

We give the results of all orbit fits, including the reduced $\chi^2$ 
and rms velocity scatter, in Table~\ref{allfits}.  Many of the attempted 
2-planet fits resulted in two Keplerians at nearly identical periods.  
These systems may have a physically plausible solution with a slightly 
worse $\chi^2$, but testing such possibilities is beyond the scope of 
this paper.  Hence, there may exist additional ``good'' 2-planet 
solutions which have been missed by our approach.  The Keplerian fitting 
methods used here incorporate no physics: they are simply seeking a 
lowest-$\chi^2$ solution regardless of the physicality of the resulting 
system parameters.  A 1:1 resonant configuration is dynamically 
possible, as evidenced by the abundance of such ``Trojan'' objects in 
our own Solar system \citep{levison97, jonti1, jonti2}.  However, the 
radial-velocity signature of a 1:1 configuration is extremely difficult 
to disentangle \citep{laughlin02,g12}.  Some extrasolar planetary 
systems have been proposed to host dynamically stable planets in 1:1 
configurations \citep{goz06, cress09, schwarz09, funk12}.  However, 
owing to the difficulty of maintaining dynamically stable 
configurations, we consider such cases to be beyond the scope of this 
work.  In this section, we will focus on those systems where a 2-planet 
fit produced a ``plausible'' result (Table~\ref{promising}) with a 
$\chi^2$ and rms similar to, or better than, the single-eccentric-planet 
fit as given in Table~\ref{allfits}.

\begin{figure}
\plotone{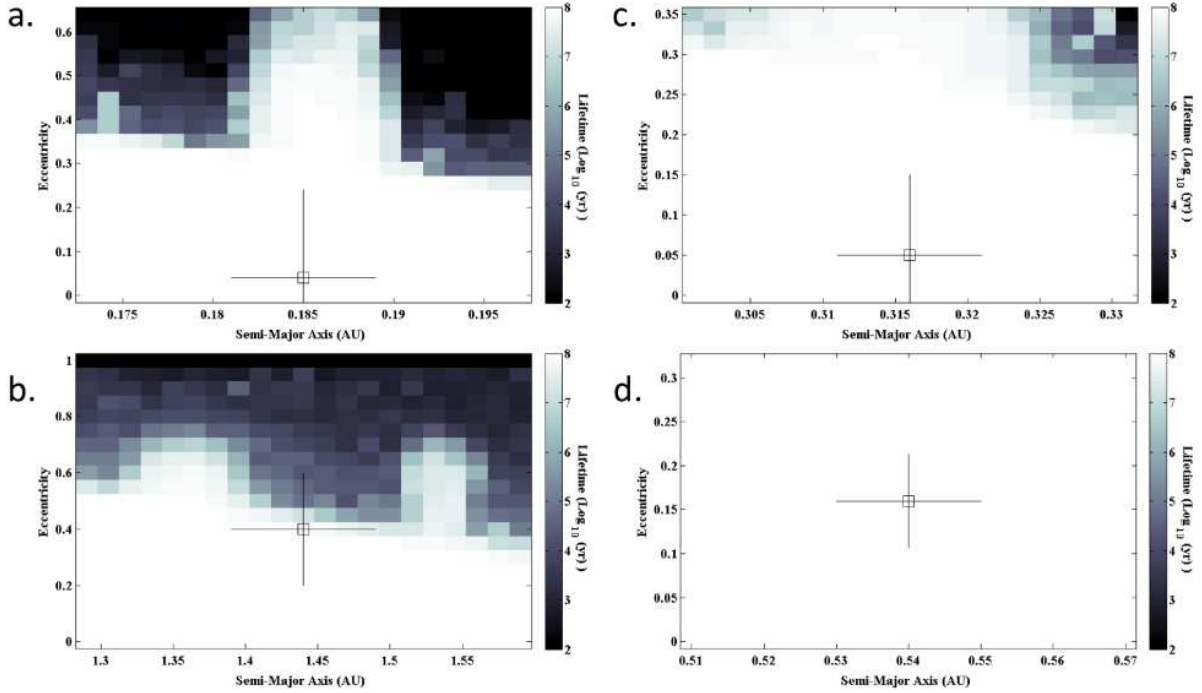}
\caption{Dynamical stability for four proposed 2-planet systems, as a 
function of the semi-major axis, $a$, and eccentricity, $e$, of planet 2 
(as given in Table~\ref{promising}.  Ranges shown are $3\sigma$ in $a$ 
and $e$ for each system.  The mean lifetime of the planetary system (in 
$log_{10}$ (lifetime/yr)) at a given $a-e$ coordinate is denoted by the 
color of the plot.  The lifetime at each $a-e$ location is the mean 
value of 25 separate integrations carried out on orbits at that $a-e$ 
position (testing a combination of 5 unique $\omega$ values, and 5 
unique $M$ values).  The nominal best-fit orbit for the outer planet is 
shown as the small open square with $\pm~1$-$\sigma$ error bars.  Panel 
(a): HD\,3651; panel (b): HD\,7449; panel (c): HD\,52265; panel (d): 
HD\,65216. }
\label{dynamics1}
\end{figure}

\textit{HD 3651} -- The best 2-planet fit resulted in a 2:1 period 
commensurability, with the second signal at $P=31.08$ days.  This is 
quite close to the monthly observing window, so the fit results in large 
uncertainties in phase ($\omega$ and mean anomaly).  The $\chi^2$ and 
rms are similar to, but slightly higher than, the one-planet fit.  
Detailed dynamical simulations (Figure~\ref{dynamics1}, panel a) clearly 
show the 2:1 resonance as a vertical strip of stability throughout the 
range of allowed eccentricities, and the best fit for a second planet 
places it comfortably within an extremely stable region.

\textit{HD 7449} -- The two-planet model results in two giant planets, 
with minimum masses of 1.2 and 0.4 \Mjup\ (Table~\ref{promising}).  The 
dynamical tests (Figure~\ref{dynamics1}, panel b) show that the best-fit 
eccentricity for the second (innermost) planet is on the edge of a 
stable region.  As the eccentricity of the second planet is 
increased beyond $e=0.4$, the system stability is quickly degraded.

\textit{HD 52265} -- As with HD\,3651, this system also gave a 2:1 
configuration, with the second planet at half the period of the known 
planet, and a slightly worse $\chi^2$ and rms.  Figure~\ref{dynamics1} 
(panel c) shows that the candidate planet is well within a broad stable 
region.

\textit{HD 65216} -- A second planet can be fit, with $P=152.6\pm$0.5 
days and a mass of 0.17$\pm$0.03 \Mjup.  The initial dynamical check in 
\textit{Systemic} showed this system to be stable, and further dynamical 
mapping (Figure~\ref{dynamics1}, panel d) shows that the entire 3-sigma 
$a-e$ parameter space is stable.

\begin{figure}
\plotone{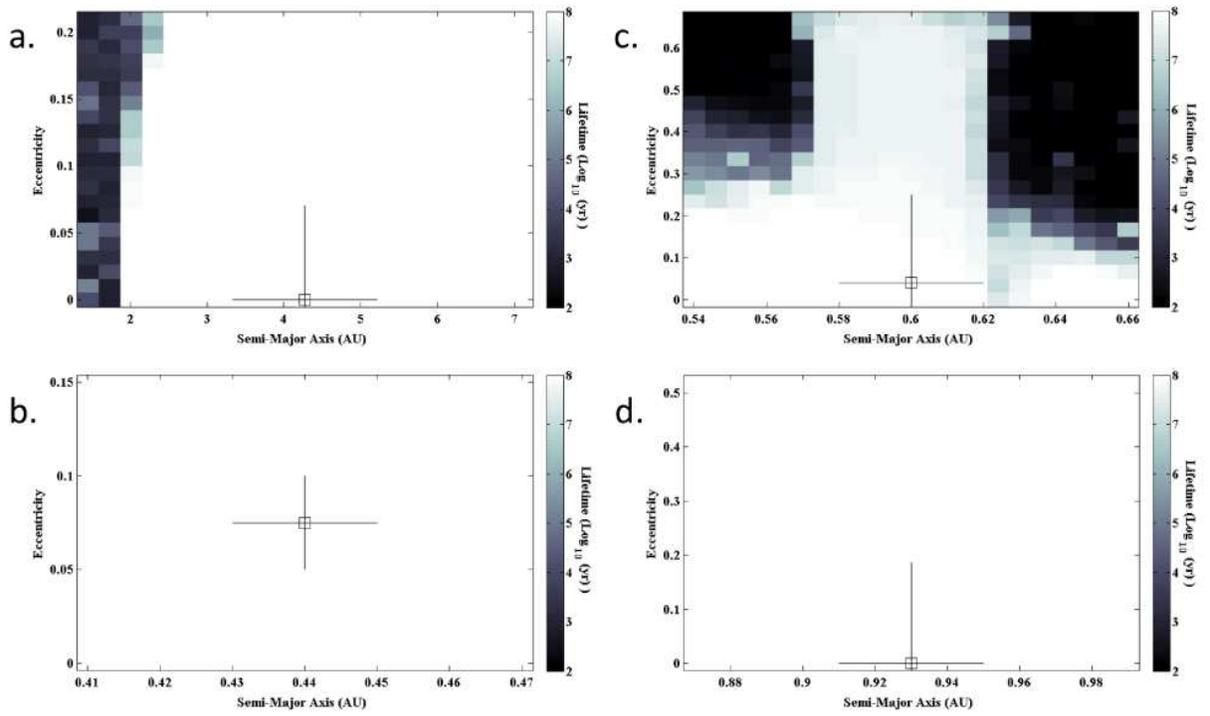}
\caption{Same as Figure~\ref{dynamics1}, but for the following systems: 
panel (a): HD\,85390; panel (b): HD\,89744; panel (c): HD\,92788; panel 
(d): HD\,117618. }
\label{dynamics2}
\end{figure}

\textit{HD 85390} -- Here, fitting a second planet improves the $\chi^2$ 
(12.2 to 4.3) and rms (2.3\,\ms to 1.4\,\ms).  The eccentricities of 
both planets are then consistent with zero.  The second planet would be 
a Jupiter analog (e.g.~Wittenmyer et al.~2011a), with $P\sim$10 yr and a 
mass of 0.20$\pm$0.02\Mjup.  The detailed dynamical map 
(Figure~\ref{dynamics2}, panel a) shows a broad swath of stability for 
all orbits with $a\gtsimeq$2\,AU.

\textit{HD 89744} -- The possible second planet has a period of 85.2 
days, very close to a 3:1 commensurability with the known 258-day 
planet.  The dynamics of the system (Figure~\ref{dynamics2}, panel b) 
show that all permissible orbits are stable, despite the relatively high 
mass of the candidate planet (3.2 \Mjup).

\textit{HD 92788} -- Figure~\ref{dynamics2} (panel c) shows this planet 
candidate is almost certainly trapped in the 2:1 resonance with the 
known 326-day planet.  This resonance would allow relatively large 
eccentricities (and even crossing orbits) to be long-term stable.

\textit{HD 117618} -- Adding a second planet improves the $\chi^2$ (8.1 
to 5.7) and rms (5.6\ms\ to 4.6\ms).  With a period of 318$\pm$2 days, 
the second signal is far enough from one year to allay fears of 
aliasing.  A dynamical map for this Saturn-mass candidate is shown in 
Figure~\ref{dynamics2} (panel d) -- again, there is a vast region of 
stability across the 3$\sigma$ parameter space.

\textit{GJ 649} -- As the two proposed planets are very widely 
separated, and of low mass, we did not perform the dynamical testing for 
this system.  The planet candidates are separated by 21.8 mutual Hill 
radii, which is certainly sufficient dynamical room for any interaction 
to be negligible \citep{chambers96}.

\textit{HD 192310 (=GJ 785)} -- While this system does not appear in 
Table~\ref{promising}, we note that the 2-planet fit shown in 
Table~\ref{allfits} suggests a Neptune-mass second planet with period 
629$\pm$64 days.  This is broadly consistent (within $\sim\,1.5\sigma$) 
with the claim of a 0.07\Mjup\ planet by \citet{pepe11}, with 
$P=526\pm$9 days.  

\citet{ang10} have also approached this problem of disentangling single 
eccentric planets from near-circular two-planet systems, specifically 
considering the case of the 2:1 resonance.  That work showed that about 
35\% of known single-planet systems were indistiguishable from 2:1 
resonant solutions.  By comparison, 9/82=11\% of the systems examined in 
the present work resulted in a 2:1 configuration (Table~2).  We also 
find that a further 24/82=29\% of systems tested resulted in co-orbital 
(1:1) configurations.  \citet{hollis12} also performed an extensive and 
self-consistent Bayesian re-analysis of available radial-velocity data 
for 94 exoplanet systems.  As part of their analysis, they attempted 
two-planet fits to the known single-planet systems (Hollis et al.~2012, 
Table~A2).  Unlike this work, their two-planet fits were not restricted 
to low eccentricities.  For example, \citet{hollis12} find a two-planet 
solution for HD\,3651 where $(P_{1},e_{1})=(62.25,0.60)$ and 
$(P_{2},e_{2})=(295,0.32)$.  By contrast, our result in 
Table~\ref{promising} gives a 2:1 configuration with $e\ltsimeq\,0.06$ 
for both planets.

\begin{figure}
\plotone{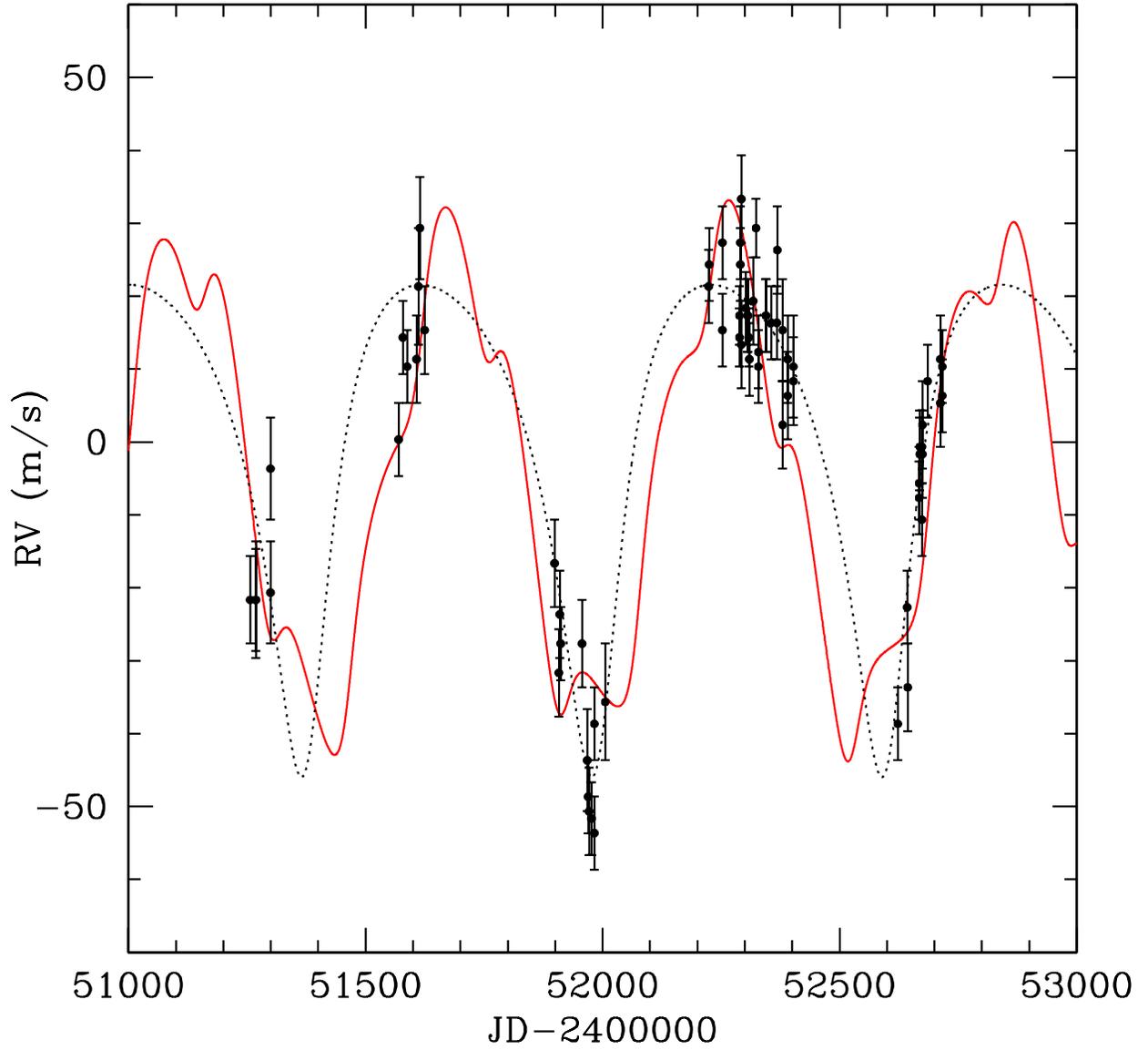}
\caption{Data for HD\,65216 \citep{mayor04} overplotted with one-planet 
(dashed line) and two-planet (solid line) models.  At present, the 
models are essentially indistinguishable, but they diverge in the future 
(Figure~\ref{orbits2}). }
\label{comparenow}
\end{figure}

\begin{figure}
\plotone{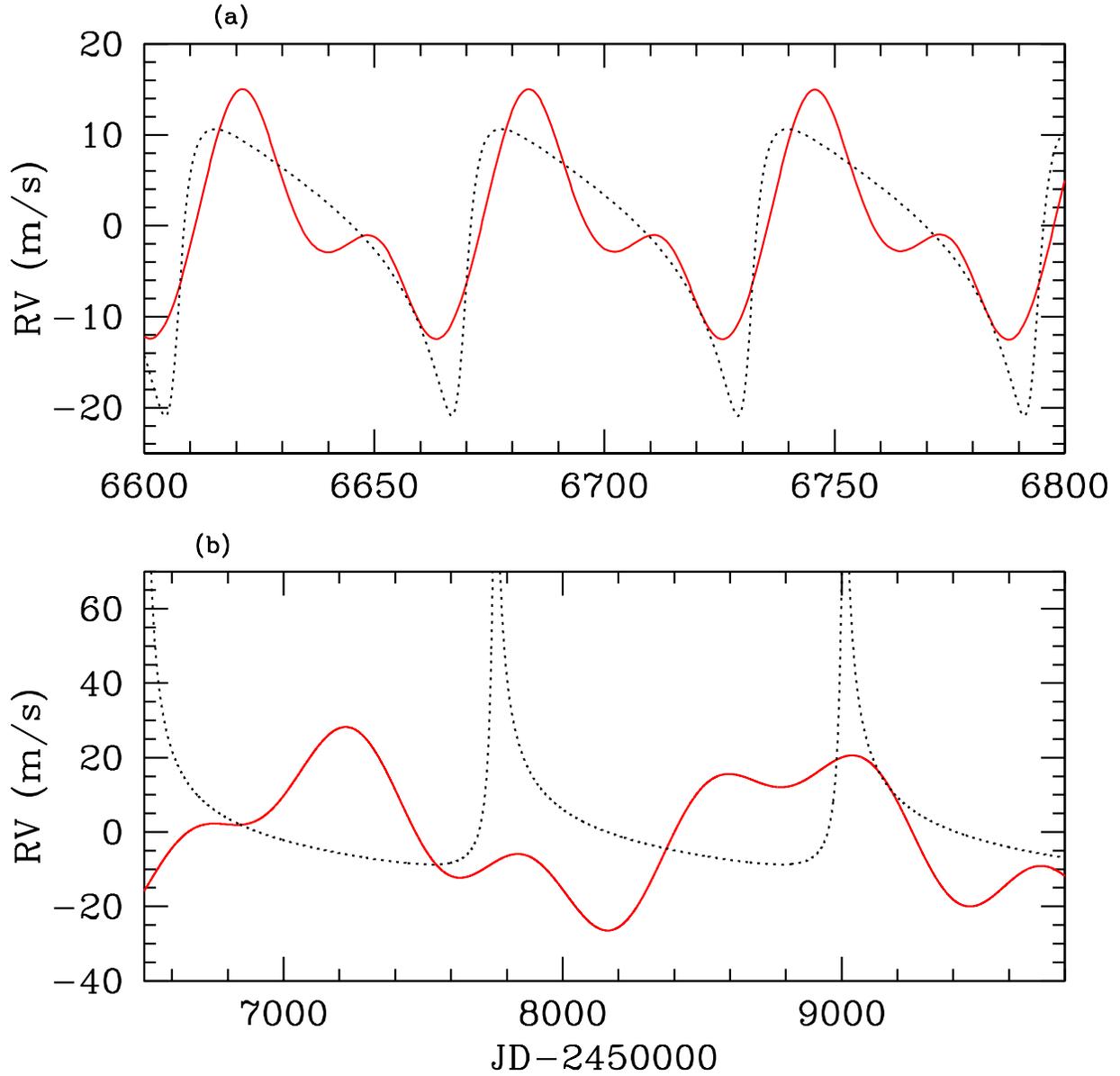}
\caption{Model orbits for single and double-planet systems.  In each 
panel, the solid (red) line is the two-planet model 
(Table~\ref{promising}), and the dashed line is the eccentric 
single-planet model.  These plots show when in the near future the two 
models could be best distinguished.  Panel (a)-- HD\,3651; panel (b)-- 
HD\,7449.}
\label{orbits1}
\end{figure}

\begin{figure}
\plotone{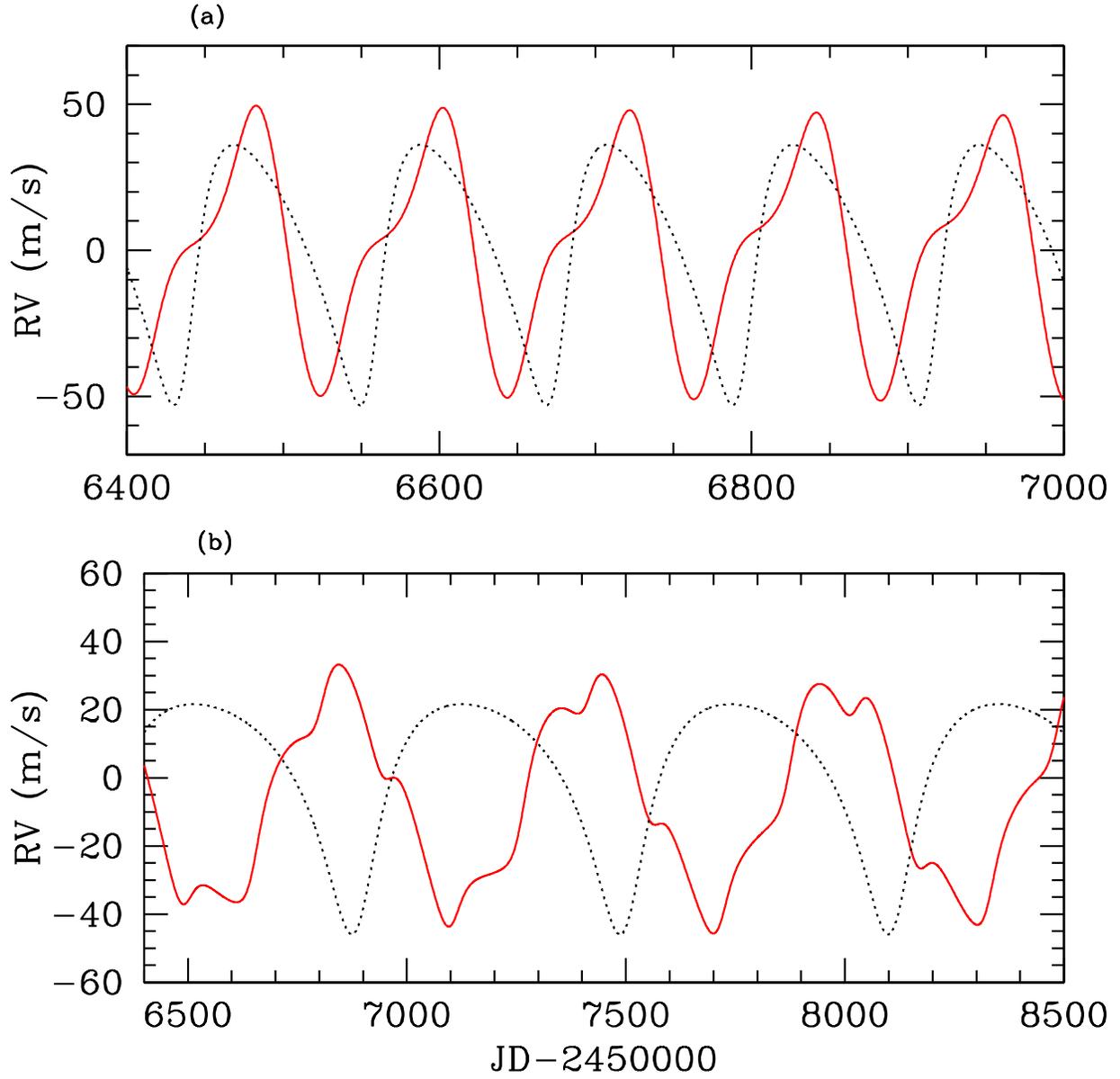}
\caption{Same as Figure~\ref{orbits1}, but for the following systems: 
panel (a)-- HD\,52265; panel (b)-- HD\,65216. }
\label{orbits2}
\end{figure}

\begin{figure}
\plotone{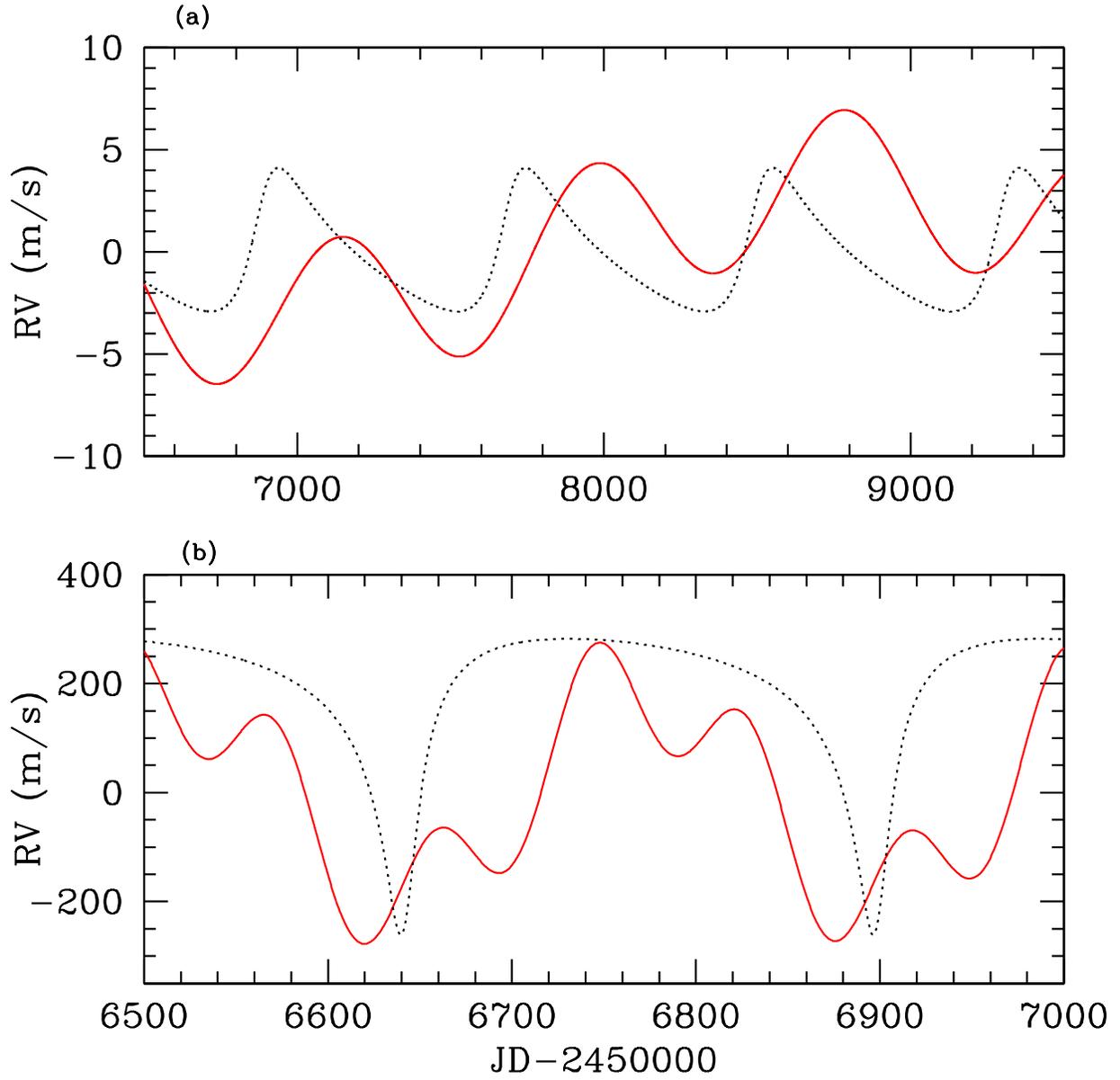}
\caption{Same as Figure~\ref{orbits1}, but for the following systems:
panel (a)-- HD\,85390; panel (b)-- HD\,89744. }
\label{orbits3}
\end{figure}

\begin{figure}
\plotone{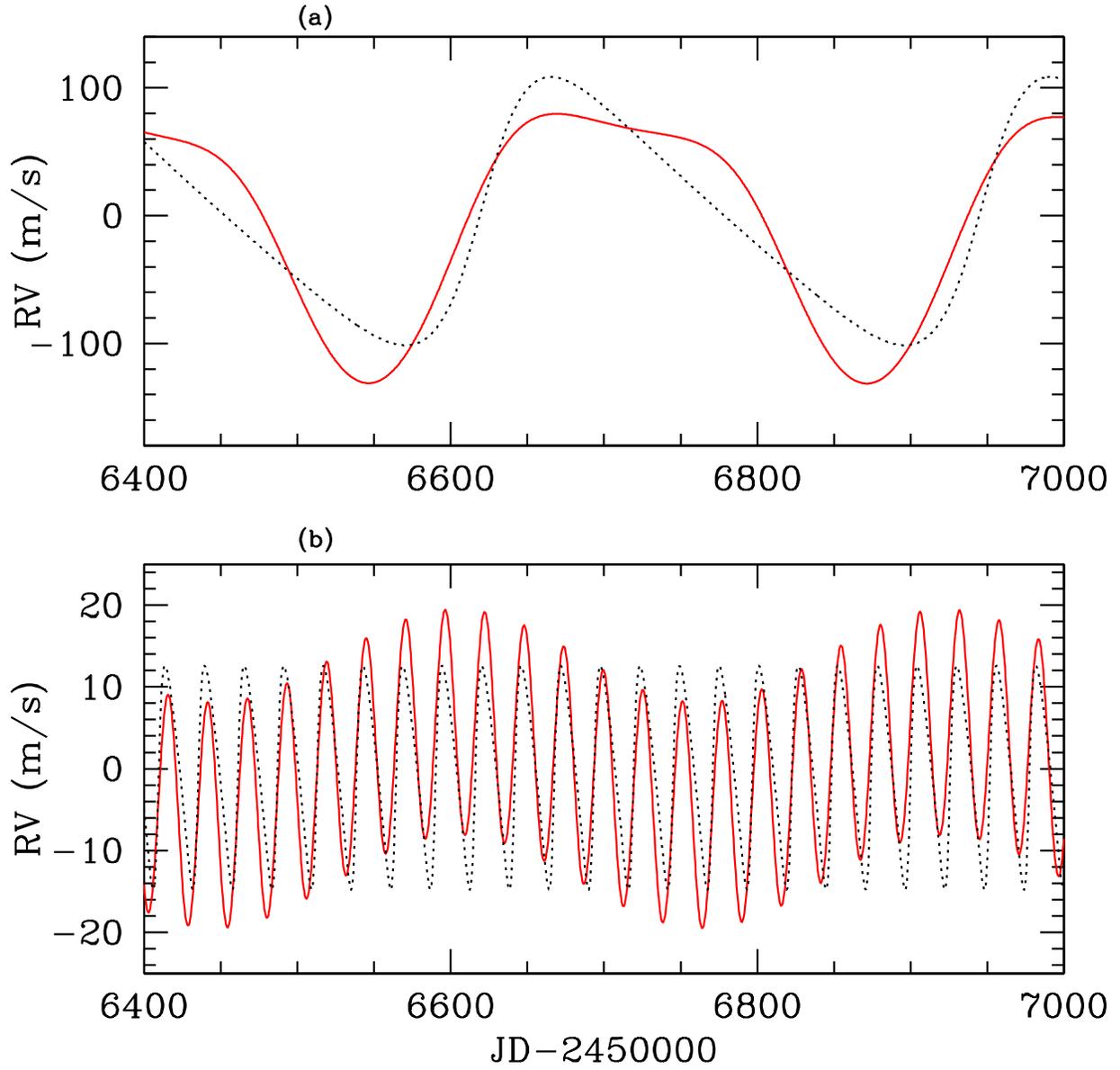}
\caption{Same as Figure~\ref{orbits1}, but for the following systems:
panel (a)-- HD\,92788; panel (b)-- HD\,117618. }
\label{orbits4}
\end{figure}

\begin{figure} 
\plotone{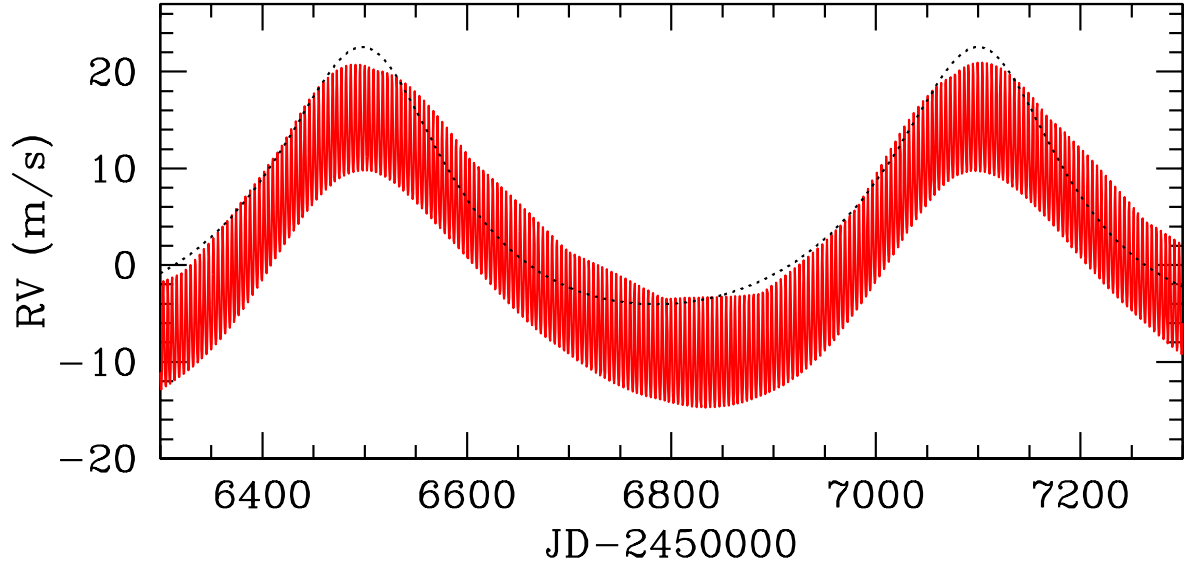} 
\caption{Same as Figure~\ref{orbits1}, but for GJ\,649.  The extremely 
short (4.5 day) period of the candidate second planet results in the 
grey region when the model velocity curve is plotted. }
\label{orbits5} 
\end{figure}

Ultimately, there is no substitute for sampling density: given infinite 
observational resources, one would ideally observe every system as often 
as possible \citep{monster}.  Such techniques have been successful in 
identifying low-mass planets \citep{16417paper, 61vir, alphacen} and in 
clarifying the true orbital period and mass of eccentric planets where 
large velocity excursions occur on short timescales \citep{37605, 
endl06}.  It is, of course, more practical to optimally plan one's 
observations to confirm suspected planet candidates \citep{scheduling}.  
All of the systems discussed above withstood detailed dynamical 
scrutiny, as shown in Figures~\ref{dynamics1} and \ref{dynamics2}.  For 
those potential 2-planet systems, we now ask how one may observationally 
discern between the one- and two-planet models.  At present, the two 
models are typically close to each other in goodness-of-fit; an example 
is shown in Figure~\ref{comparenow} for HD\,65216.  In the future, 
however, the two possible models may diverge.  In 
Figures~\ref{orbits1}--\ref{orbits5}, we overplot the two models for 
each system, to give a qualitative estimate of the optimal times to 
observe them.  When they diverge sufficiently, well-timed observations 
of reasonable precision ($\sim$2-3 \ms) can distinguish between the 1- 
and 2-planet models.  We note, however, that the 2-planet fits often 
have large phase gaps (and hence relatively large uncertainties in $P$ 
and $T_0$) which can shift the model curves shown here.  For this 
reason, we leave more detailed quantitative simulations of detectability 
(e.g. Wittenmyer et al. 2013) for future work.

\section{Summary and Conclusions}

We have examined 82 known moderately eccentric single-planet systems, 
reanalysing the available radial-velocity data to test the hypothesis 
that some may actually be low-eccentricity two-planet systems.  We have 
identified nine particularly promising candidate systems, and performed 
detailed dynamical stability simulations of the candidate planets.  All 
of the systems proved to be dynamically stable on timescales of at least 
$10^8$ yr.  We have also given model orbits to show qualitatively when 
the one- and two-planet solutions diverge enough to be distinguishable 
by future well-timed radial-velocity observations.  Our results suggest 
that at least 11\% of apparently single-planet systems may in fact host 
two low-eccentricity planets, a figure likely to rise as more 
observations are obtained.  The difference between the one- and 
two-planet solutions should typically be observable within the next 3 
years, adding weight to the case for continued observations of these 
systems.


\acknowledgements

We gratefully acknowledge the UK and Australian government support of 
the Anglo-Australian Telescope through their PPARC, STFC and DIISR 
funding; STFC grant PP/C000552/1, ARC Grant DP0774000 and travel support 
from the Australian Astronomical Observatory.  The work was supported by 
iVEC through the use of advanced computing resources located at the 
Murdoch University, in Western Australia.  RW is grateful to Nanjing 
University for the support of his visit to Nanjing.  S.~Wang's visit to 
UNSW was funded by Faculty Research Grant PS27262.

This research has made use of NASA's Astrophysics Data System (ADS), and 
the SIMBAD database, operated at CDS, Strasbourg, France.  This research 
has made use of the Exoplanet Orbit Database and the Exoplanet Data 
Explorer at exoplanets.org \citep{wright11}.



\begin{deluxetable}{lll}
\tabletypesize{\scriptsize}
\tablecolumns{3}
\tablewidth{0pt}
\tablecaption{Summary of Radial-Velocity Data }
\tablehead{
\colhead{Star} & \colhead{$N$} & \colhead{Source}
}
\startdata
\label{datasummary}
HD 1237 & 61 & \citet{naef01} \\
HD 1690 & 41 & \citet{moutou11} \\
HD 2039 & 46 & \citet{tinney03}\tablenotemark{a} \\
HD 3651 & 163 & \citet{butler06} \\
HD 3651 & 35 & \citet{witt09} \\
HD 4113 & 130 & \citet{tamuz08}  \\
HD 4203 & 23 & \citet{butler06} \\
HD 5388 & 68 & \citet{santos10} \\
HD 7449 & 82 & \citet{dum11} \\
HD 8574 & 41 & \citet{perrier03} \\
HD 8574 & 60 & \citet{witt09} \\
HD 8574 & 26 & \citet{butler06} \\
HD 11506 & 26 & \citet{fischer07} \\ 
HD 16175 & 44 & \citet{peek09} \\
HD 20782 & 47 & \citet{jones06}\tablenotemark{a} \\
HD 20868 & 48 & \citet{moutou09} \\
HD 22781 & 32 & \citet{diaz12} \\
HD 23127 & 44 & \citet{otoole07}\tablenotemark{a} \\
HD 30562 & 45 & \citet{fischer09} \\
HD 31253 & 39 & \citet{mes11} \\
HD 33283 & 25 & \citet{johnson06} \\
HD 33636 & 32 & \citet{vogt02} \\
HD 38283 & 61 & \citet{tinney11}\tablenotemark{a} \\
HD 39091 & 67 & \citet{jones02}\tablenotemark{a} \\
HD 45350 & 73 & \citet{endl06} \\
HD 45350 & 30 & \citet{marcy05} \\
HD 45652 & 45 & \citet{santos08} \\
HD 52265 & 91 & \citet{naef01} \\
HD 52265 & 28 & \citet{butler06} \\
HD 65216 & 70 & \citet{mayor04} \\
HD 81040 & 26 & \citet{soz06} \\
HD 85390 & 58 & \citet{mordasini11} \\
HD 86264 & 37 & \citet{fischer09} \\
HD 87883 & 69 & \citet{fischer09} \\
HD 89744 & 50 & \citet{butler06} \\
HD 89744 & 42 & \citet{witt09} \\
HD 90156 & 66 & \citet{mordasini11} \\
HD 92788 & 55 & \citet{mayor04} \\
HD 92788 & 58 & \citet{butler06} \\
HD 96127 & 50 & \citet{gettel12} \\
HD 96167 & 47 & \citet{peek09} \\
HD 99706 & 24 & \citet{johnson11} \\
HD 100777 & 29 & \citet{naef07} \\
HD 102365 & 168 & \citet{tinney11a}\tablenotemark{a} \\
HD 106252 & 40 & \citet{perrier03} \\
HD 106252 & 70 & \citet{witt09} \\
HD 106270 & 20 & \citet{johnson11} \\
HD 108147 & 57 & AAT\tablenotemark{a} \\
HD 108147 & 118 & \citet{pepe02} \\
HD 117618 & 70 & \citet{tinney05}\tablenotemark{a} \\
HD 118203 & 43 & \citet{dasilva06} \\
HD 126614A & 70 & \citet{howard10a} \\
HD 131664 & 41 & \citet{moutou09} \\
HD 132406 & 21 & \citet{dasilva07} \\
HD 136118 & 37 & \citet{butler06} \\ 
HD 136118 & 68 & \citet{witt09} \\
HD 137388 & 62 & \citet{dum11} \\
HD 137510 & 76 & \citet{endl04} \\
HD 137510 & 13 & \citet{diaz12} \\ 
HD 141937 & 81 & \citet{udry02} \\
HD 142022 & 76 & \citet{egg06} \\
HD 142415 & 137 & \citet{mayor04} \\
HD 142415 & 22 & AAT\tablenotemark{a} \\
HD 145377 & 64 & \citet{moutou09} \\
HD 153950 & 49 & \citet{moutou09} \\
HD 154672 & 6 & \citet{jenkins09} \\
HD 154672 & 16 & \citet{lopez08} \\
HD 156279 & 15 & \citet{diaz12} \\
HD 156846 & 54 & \citet{tamuz08} \\
HD 171028 & 19 & \citet{santos11} \\
HD 171238 & 99 & \citet{seg10} \\
HD 175541 & 29 & \citet{johnson07} \\
HD 175167 & 13 & \citet{arr10} \\
HD 187085 & 64 & \citet{jones06}\tablenotemark{a} \\
HD 190228 & 51 & \citet{perrier03} \\
HD 190228 & 50 & \citet{witt09} \\
HD 196885 & 76 & \citet{fischer09} \\
HD 196885 & 102 & \citet{corr08} \\
HD 204941 & 35 & \citet{dum11} \\
HD 210277 & 69 & \citet{butler06} \\
HD 210277 & 21 & \citet{witt07} \\
HD 210277 & 42 & \citet{naef01} \\
HD 213240 & 72 & \citet{santos01} \\
HD 213240 & 35 & AAT\tablenotemark{a} \\
HD 217786 & 17 & \citet{moutou11} \\
HD 216437 & 50 & \citet{jones02}\tablenotemark{a} \\
HD 216437 & 21 & \citet{mayor04} \\
HD 216770 & 16 & \citet{mayor04} \\
HD 218566 & 56 & \citet{mes11} \\
HD 222582 & 37 & \citet{butler06} \\
HD 240237 & 40 & \citet{gettel12} \\
HIP 2247 & 26 & \citet{moutou09} \\
HIP 5158 & 54 & \citet{locurto10} \\
iota Dra & 119 & \citet{butler06} \\
iota Dra & 56 & \citet{frink02} \\
GJ 649 & 43 & \citet{johnson10} \\
GJ 785 & 75 & \citet{howard11} \\
HIP 57050 & 37 & \citet{hag10} \\
GJ 676A & 69 & \citet{for11} \\
14 Her & 49 & \citet{butler06} \\
14 Her & 35 & \citet{witt07} \\
14 Her & 119 & \citet{naef04} \\
42 Dra & 45 & \citet{dol09} \\
70 Vir & 74 & \citet{butler06} \\
70 Vir & 35 & \citet{naef04} \\
\enddata
\tablenotetext{a}{Includes additional unpublished AAPS data, given in 
Tables A1-A12 in the Appendix.}
\end{deluxetable}


\begin{deluxetable}{llllllllllllll}
\rotate
\tabletypesize{\scriptsize}
\tablecolumns{14}
\tablewidth{0pt}
\tablecaption{Summary of Results }
\tablehead{
\colhead{Star} & \colhead{Method} & \colhead{$\chi^{2}_{\nu}$} & 
\colhead{RMS} & \colhead{$P_1$} & \colhead{$m_1$ sin $i$} & 
\colhead{$M_1$} & \colhead{$e_1$} & \colhead{$\omega_1$} & 
\colhead{$P_2$} & \colhead{$m_2$ sin $i$} & \colhead{$M_2$} & \colhead{$e_2$} 
& \colhead{$\omega_2$}\\
\colhead{} & \colhead{} & \colhead{} & \colhead{(\ms)} & 
\colhead{(days)} & \colhead{(\Mjup)} & \colhead{(degrees)} & \colhead{} 
& \colhead{degrees} & \colhead{(days)} & \colhead{(\Mjup)} & 
\colhead{(degrees)} & \colhead{} & \colhead{(degrees)} \\
}
\startdata
\label{allfits}
HD 1237 & 1\tablenotemark{a} & 3.40 & 18.9 & 133.7(2) & 3.4(2) & 
317(2) & 0.51(2) & 291(3) & & & & & \\ 
        & 2\tablenotemark{b} & 10.06 & 31.1 & 117.6(6.6) & 4(12) & 
140(32) & 0.18(10) & 16(48) & 106.9(10.2) & 4(11) & 71(26) & 0.2(0.2) & 
192(52) \\
HD 1690 & 1 & 359.7 & 35.3 & 527(2) & 4.7(5) & 77(5) & 0.74(11) & 
103(13) & & & & & \\
        & 2 & 353.5 & 30.35 & 494(11) & 8(8) & 265(33) & 0.06(23) & 
243(44) & 401.3(8.8) & 5(4) & 302(42) & 0.2(0.2) & 328(37) \\
HD 2039 & 1 & 6.43 & 13.7 & 1110(3) & 4.5(1.1) & 57(7) & 0.64(6) & 
342(3) & & & & & \\
        & 2 & 35.83 & 27.3 & 1075(49) & 6(7) & 338(57) & 0.02(31) & 
304(38) & 1095(19) & 7(7) & 13(55) & 0.19(16) & 70(35) \\
HD 3651 & 1 & 3.82 & 6.3 & 62.22(1) & 0.23(1) & 121(9) & 0.60(4) & 
243(5) & & & & & \\
        & 2 & 5.68 & 6.8 & 62.20(0.04) & 0.22(3) & 241(33) & 0.03(33) & 
112(21) & 31.07(2) & 0.09(5) & 78(65) & 0.0(2) & 328(35) \\
HD 4113 & 1 & 4.28 & 9.5 & 526.61(8) & 1.66(7) & 217.1(3) & 0.899(6) & 
320(2) & & & & & \\
        & 2 & 215.83 & 71.0 & 506(14) & 3.1(1.6) & 15(16) & 0.00(38) & 
111(34) & 404.6(2.2) & 2.7(2.1) & 314(46) & 0.18(23) & 268(20) \\
HD 4203 & 1 & 4.14 & 5.8 & 434(2) & 2.2(7) & 226(14) & 0.7(1) & 346(5) 
& & & & & \\
        & 2 & 35.66 & 12.9 & 422.6(11.3) & 2.6(2.0) & 156(53) & 0.15(21) 
& 94(70) & 433(22) & 2.6(2.0) & 259(41) & 0.20(35) & 219(65) \\
HD 5388 & 1 & 2.64 & 4.1 & 777.2(3.5) & 2.0(1) & 284(6) & 0.40(2) & 
324(4) & & & & & \\
        & 2 & 4.07 & 4.8 & 776(39) & 2.4(1.7) & 255(21) & 0.16(20) & 
52(30) & 769.3(30.1) & 2.2(1.6) & 317(18) & 0.18(20) & 208(23) \\
HD 7449 & 1 & 47.58 & 4.2 & 1251(17) & 1.6(7) & 56(5) & 0.848 & 
337(3) & & & & & \\
        & 2 & 71.15 & 5.8 & 1494(138) & 3.2(3.0) & 82(32) & 0.2(9) & 
5(57) & 1738(1465) & 2.8(3.4) & 65(46) & 0.1(3) & 254(40) \\
HD 8574 & 1 & 2.21 & 14.0 & 227.0(2) & 1.81(8) & 36(6) & 0.30(3) & 
27(5) & & & & & \\
        & 2 & 2.33 & 15.0 & 227.4(2) & 1.98(7) & 56(46) & 0.12(4) & 
11(66) & 17.75(2) & 0.13(10) & 105(35) & 0.1(3) & 190(30) \\
HD 11506 & 1 & 15.88 & 9.79 & 1436(102) & 5.0(7) & 194(17) & 0.43(17) &
270(9) & & & & & \\
         & 2 & 9.9 & 6.6 & 1333(683) & 4.7(1.1) & 220(22) & 0.1(2) & 
245(51) & 370 & 1(6) & 41(33) & 0.2(2) & 221(50) \\
HD 16175 & 1 & 2.36 & 8.6 & 990(9) & 4.4(3) & 189(4) & 0.60(3) & 221(3) 
& & & & & \\
         & 2 & 5.83 & 13.7 & 1087(32) & 9(2) & 302(14) & 0.02(5) & 279 & 
1026(19) & 12(2) & 234(18) & 0.2 & 163 \\
HD 20782 & 1 & 6.56 & 6.5 & 597.08(6) & 1.35(9) & 340.5(2) & 0.960 & 
140(2) & & & & & \\
         & 2 & 385.25 & 36.2 & 586(27) & 5(2) & 356(30) & 0.1(1) & 
16(36) & 595(82) & 4.6(3.1) & 281(25) & 0.19(8) & 279(57) \\
HD 20868 & 1 & 1.63 & 1.8 & 380.85(9) & 1.99(9) & 15.7(3) & 0.755(2) 
& 356.2(4) & & & & & \\
         & 2 & 250.83 & 21.4 & 409.8(9.4) & 3.0(1.3) & 135(35) & 
0.17(21) & 307(31) & 50.4(2.3) & 1.0(1.2) & 286(56) & 0.12(28) & 103(29) 
\\
HD 22781 & 1 & 9.54 & 13.0 & 528.1(2) & 13.9(7) & 173.3(2) & 0.819(3) & 
316.5(8) & & & & & \\
         & 2 & 1002.75 & 121.1 & 500(18) & 23(18) & 147(39) & 0.1(3) & 
291(40) & 105(4) & 7(7) & 65(41) & 0.2(2) & 33(46) \\
HD 23127 & 1 & 8.36 & 11.7 & 1237(14) & 1.5(1) & 324(16) & 0.37(7) &
197(13) & & & & & \\
         & 2 & 19.55 & 13.5 & 1233.6(17.3) & 1.8(0.3) & 313(57) & 
0.15(10) & 208(78) & 9.91(13) & 0.08(17) & 216(50) & 0.04(30) & 149(41) 
\\
HD 30562 & 1 & 2.49 & 7.1 & 1159(16) & 1.35(6) & 326(21) & 0.76(3) & 
79(6) & & & & & \\
         & 2 & 8.17 & 12.8 & 1131.4(24.4) & 2.5(1.4) & 314(59) & 
0.16(10) & 95(28) & 825(756) & 2.0(1.6) & 290(24) & 0.20(9) & 321(48) \\
HD 31253 & 1 & 8.87 & 4.3 & 465.4(1.8) & 0.50(5) & 138(16) & 0.34(10) &
244(16) & & & & & \\
         & 2 & 6.47 & 3.4 & 463.4(3.2) & 0.51(8) & 134(55) & 0.08(11) & 
248(42) & 686.0(15.9) & 0.3(5) & 4(43) & 0.05(29) & 128(47) \\
HD 33283 & 1 & 0.64 & 3.2 & 18.179(6) & 0.33(2) & 305(4) & 0.48(4) &
156(7) & & & & & \\
         & 2 & 0.56 & 2.6 & 18.12(4) & 0.37(9) & 243(27) & 0.1(1) & 
230(65) & 47.6(3) & 0.4(2) & 32(10) & 0.2(2) & 11(73) \\
HD 33636 & 1 & 2.13 & 8.7 & 1552(135) & 7.8(5) & 276(8) & 0.39(3) &
335(5) & & & & & \\
         & 2 & 19.01 & 16.0 & 2204(613) & 9.5(3.9) & 153(63) & 0.09(16) 
& 121(31) & 970(136) & 4.3(3.1) & 195(58) & 0.20(19) & 18(48) \\
HD 38283 & 1 & 7.90 & 5.6 & 360.4(9) & 0.5(2) & 150(11) & 0.64(26) &
57(21) & & & & & \\
         & 2 & 6.73 & 4.8 & 354.5(3.7) & 0.6(2.6) & 66(43) & 0.16(14) & 
127(51) & 364.5(4.2) & 0.4(2.5) & 357(42) & 0.17(24) & 320(59) \\
HD 39091 & 1 & 11.3 & 6.2 & 2088(3) & 9.7(3) & 137(1) & 0.643(5) & 
331.5(7) & & & & & \\
         & 2 & 685.05 & 43.8 & 2093(33) & 7.5(8) & 356(8) & 0.002 & 110 
& 1056(11) & 5.0(5) & 295(18) & 0.2 & 314 \\
HD 45350 & 1 & 1.50 & 8.0 & 964(3) & 1.8(1) & 14(3) & 0.778(9) & 343(2) 
& & & & & \\
         & 2 & 12.86 & 16.9 & 989(23) & 3.4(7.8) & 225(27) & 0.17(6) & 
99(43) & 953(40) & 3.7(7.8) & 272(24) & 0.17(13) & 232(34) \\
HD 45652 & 1 & 2.88 & 13.6 & 43.7(1) & 0.47(4) & 87(13) & 0.45(6) & 
249(11) & & & & & \\
         & 2 & 3.71 & 12.6 & 43.83(24) & 0.6(1) & 66(43) & 0.17(11) & 
281(16) & 95.6(4.2) & 0.3(3) & 9(45) & 0.17(22) & 7(38) \\
HD 52265 & 1 & 2.02 & 10.9 & 119.31(78) & 1.12(6) & 24(5) & 0.35(3) & 
232(6) & & & & & \\ 
         & 2 & 2.31 & 10.8 & 119.38(25) & 1.31(7) & 19(64) & 0.19(5) & 
243(39) & 179.1(4.0) & 0.33(24) & 317(55) & 0.06(22) & 168(53) \\
HD 65216 & 1 & 1.73 & 7.1 & 612(10) & 1.22(7) & 289(17) & 0.41(6) & 
198(7) & & & & & \\
         & 2 & 1.57 & 6.5 & 574.2(7.1) & 1.4(2) & 3(66) & 0.15(8) & 
82(34) & 270.7(3.3) & 0.4(2) & 294(59) & 0.02(9) & 298(22) \\
HD 81040 & 1 & 5.68 & 27.9 & 1005(10) & 6.7(4) & 286(37) & 0.59(4) & 
85(4) & & & & & \\
         & 2 & 4.02 & 22.8 & 1091(23) & 9(4) & 15(69) & 0.18(17) & 
51(27) & 262.2(6.9) & 2(3) & 250(24) & 0.20(29) & 19(49) \\
HD 85390 & 1 & 12.22 & 2.3 & 806(19) & 0.11(1) & 62(39) & 
0.59(fixed) & 301(10) & & & & & \\
         & 2 & 5.56 & 1.6 & 799(34) & 0.14(1) & 22(21) & 0.01(13) & 
307(18) & 2491(5399) & 0.18(15) & 305(39) & 0.03(26) & 151(42) \\
HD 86264 & 1 & 2.76 & 26.9 & 1520(34) & 6.8(4) & 284(37) & 0.82(17) & 
296(27) & & & & & \\
         & 2 & 3.2 & 30.9 & 1416(66) & 9(4) & 276(29) & 0.1(2) & 310(22) 
& 194(12) & 2(2) & 119(51) & 0.16(30) & 318(42) \\
HD 87883 & 1 & 4.16 & 8.9 & 2762(11) & 1.8(2) & 2(7) & 0.55(17) & 
290(16) & & & & & \\
         & 2 & 3.88 & 8.6 & 2934(113) & 2.1(3) & 44(6) & 0.17 & 283 & 
342(2) & 0.3(1) & 95(15) & 0.20 & 305 \\
HD 89744 & 1 & 2.58 & 15.2 & 256.78(5) & 8.5(3) & 321.5(5) & 0.673(7) & 
195(1) & & & & & \\
         & 2 & 58.5 & 72.3 & 256.7(1) & 11.9(1.3) & 325(51) & 0.20(2) & 
189(48) & 256(344) & 5(2) & 201(53) & 0.2(3) & 118(53) \\
HD 90156 & 1 & 9.99 & 1.2 & 49.79(6) & 0.055(5) & 218(13) & 0.34(6) &
112(12) & & & & & \\
         & 2 & 8.56 & 1.1 & 49.65(7) & 0.069(5) & 207(60) & 0.18(10) & 
98(37) & 13.52(4) & 0.015(14) & 292(37) & 0.03(29) & 19(62) \\
HD 92788 & 1 & 2.29 & 8.6 & 325.7(2) & 3.5(1) & 80(3) & 0.336(9) &
276(2) & & & & & \\
         & 2 & 4.04 & 10.3 & 325(3) & 4(1) & 44(53) & 0.16(15) & 0(58) & 
327(6) & 3(2) & 132(47) & 0.2(2) & 152(45) \\
HD 96127 & 1 & 77.56 & 49.2 & 636(19) & 4.1(9) & 191(39) & 0.36(13) &
155(7) & & & & & \\
         & 2 & 39.53 & 40.0 & 652(16) & 3.7(5) & 227(47) & 0.18 & 128 & 
5.573(2) & 0.50(9) & 227(25) & 0.10 & 228 \\
HD 96167 & 1 & 11.32 & 4.4 & 498(2) & 0.68(4) & 329(6) & 0.71(6) & 288(9) 
& & & & & \\
         & 2 & 20.96 & 6.4 & 518 & 1(1) & 283(54) & 0.0(0.4) & 153(51) & 
508(8) & 1.4(1.0) & 339(52) & 0.2(2) & 268(28) \\
HD 99706 & 1 & 23.57 & 5.7 & 812(22) & 1.7(1) & 16(17) & 0.31(9) & 
357(17) & & & & & \\
         & 2 & 16.8 & 3.9 & 853(71) & 1.8(9) & 246(25) & 0.2(2) & 
159(62) & 379(19) & 0.6(1.0) & 111(46) & 0.1(3) & 249(32) \\
HD 100777 & 1 & 1.49 & 1.8 & 383.7(1.1) & 1.16(6) & 352(2) & 0.358(18) & 
203(3) & & & & & \\
          & 2 & 1.86 & 1.7 & 382.8(3.6) & 1.3(7) & 33(46) & 0.16(14) & 
155(50) & 127.3(2.9) & 0.18(16) & 308(34) & 0.12(26) & 199(39) \\
HD 102365 & 1 & 10.70 & 3.8 & 122.1(4) & 0.048(8) & 322(31) & 0.17(16) &
56(fixed) & & & & & \\
          & 2 & 9.31 & 3.5 & 122.18(73) & 0.07(3) & 276(61) & 0.07(21) & 
105(51) & 927(35) & 0.1(2) & 233(56) & 0.00(36) & 44(53) \\
HD 106252 & 1 & 1.42 & 12.2 & 1531(5) & 7.0(3) & 41(2) & 0.48(1) & 293(2)
& & & & & \\
          & 2 & 3.34 & 17.5 & 1510(12) & 6.8(2) & 26(38) & 0.07(3) & 
303(22) & 760 & 2(2) & 44(32) & 0.2(2) & 348(41) \\
HD 106270 & 1 & 30.1 & 8.5 & 2658(880) & 11(2) & 277(27) & 0.36(15) & 
16(6) & & & & & \\
          & 2 & 49.38 & 8.5 & 2539(271) & 12(4) & 302(28) & 0.14(15) & 
345(55) & 885(82) & 1.3(2.3) & 83(51) & 0.2(3) & 26(42) \\
HD 108147 & 1 & 6.06 & 15.4 & 10.9013(7) & 0.31(2) & 60(4) & 0.53(4) & 
307(5) & & & & & \\
          & 2 & 7.58 & 16.4 & 10.902(1) & 0.34(2) & 71(31) & 0.17(6) & 
300(38) & 91.1(9) & 0.2(2) & 118(30) & 0.04(32) & 36(28) \\
HD 117618 & 1 & 8.11 & 5.6 & 25.815(6) & 0.20(2) & 334(18) & 0.33(9) &
256(15) & & & & & \\
          & 2 & 5.88 & 4.7 & 25.81(1) & 0.21(2) & 323(64) & 0.08(10) & 
256(46) & 319.1(4.2) & 0.2(2) & 283(60) & 0.00(32) & 32(25) \\
HD 118203 & 1 & 2.26 & 19.6 & 6.1345(9) & 2.13(9) & 149(4) & 
0.31(2) & 155.7(3.6) & & & & & \\
          & 2 & 6.72 & 31.2 & 6.135(2) & 2.31(6) & 152(59) & 0.145(35) & 
154(63) & 29.0(1.0) & 0.5(4) & 168(29) & 0.18(25) & 336(54) \\
HD 126614 & 1 & 8.87 & 3.8 & 1245(12) & 0.39(3) & 328(11) & 
0.43(9) & 241(14) & & & & & \\
          & 2 & 11.03 & 4.3 & 1212(24) & 0.4(2) & 218(33) & 0.04(20) & 
340(28) & 339(18) & 0.1(2) & 192(30) & 0.0(3) & 288(43) \\
HD 131664 & 1 & 6.06 & 5.7 & 1951(42) & 18(1) & 200(9) & 0.64(2) & 
150(1) & & & & & \\
          & 2 & 347.83 & 27.7 & 1825(17) & 24(3) & 176(87) & 0.16(8) & 
189(77) & 1584(116) & 15(4) & 243(49) & 0.20(13) & 307(48) \\
HD 132406 & 1 & 1.74 & 13.2 & 975(50) & 5.6(1.6) & 241(22) & 0.34(12) &
214(24) & & & & & \\
          & 2 & 2.00 & 11.1 & 950(124) & 5.5(2.6) & 264(35) & 0.04(18) & 
192(54) & 29.2(2) & 0.5(5) & 171(42) & 0.07(34) & 120(48) \\
HD 136118 & 1 & 1.82 & 16.5 & 1187.3(2.4) & 11.7(4) & 61(3) & 
0.338(15) & 320(2) & & & & & \\
          & 2 & 2.76 & 20.0 & 1190.1(21.2) & 13(4) & 66(13) & 0.198(40) 
& 310(45) & 1291.6(9999) & 2(4) & 51(37) & 0.04(36) & 157(38) \\
HD 137388 & 1 & 17.05 & 3.2 & 355.6(2.6) & 0.32(4) & 318(33) & 
0.13(9) & 269(31) & & & & & \\
          & 2 & 13.16 & 2.5 & 330.7(4.4) & 0.3(2) & 43(51) & 0.17(15) & 
60.1(9.6) & 2436(2134) & 0.5(6) & 329(57) & 0.16(13) & 139(41) \\
HD 137510 & 1 & 5.24 & 20.4 & 800.9(5) & 26.4(1.2) & 37(1) &
0.399(8) & 32(1) & & & & & \\
          & 2 & 17.64 & 38.1 & 802(4) & 30(14) & 47(53) & 0.19(5) & 
11(68) & 804(54) & 10(14) & 315(34) & 0.2(9) & 245(39) \\
HD 141937 & 1 & 2.69 & 9.6 & 653(2) & 9.4(6) & 41(2) &
0.41(2) & 187.7(1.3) & & & & & \\
          & 2 & 4.33 & 11.6 & 659.5(8.7) & 11(1) & 54(45) & 0.183(23) & 
166(7) & 668(34) & 3.9(5) & 300(30) & 0.17(24) & 74(74) \\
HD 142022 & 1 & 1.54 & 10.4 & 1931(35) & 4.3(1.0) & 79(9) & 0.52(8) & 
169(5) & & & & & \\
          & 2 & 2.05 & 10.7 & 1894(41) & 3(2) & 260(24) & 0.01(28) & 
342(47) & 946(31) & 1(2) & 32(23) & 0.17(17) & 296(40) \\
HD 142415 & 1 & 15.03 & 14.8 & 406.6(9) & 1.8(1) & 104(3) & 0.64(2) & 
222(4) & & & & & \\
          & 2 & 19.82 & 17.0 & 407.7(4.3) & 2.2(2.7) & 182(63) & 
0.19(13) & 258(37) & 397(48) & 1.2(3.0) & 263(42) & 0.16(26) & 328(47)\\
HD 145377 & 1 & 104.0 & 16.8 & 103.96(17) & 5.8(2) & 176(5) & 
0.307(17) & 138(3) & & & & & \\
          & 2 & 92.50 & 17.2 & 103.31(15) & 5.7(1) & 147(69) & 0.074(35) 
& 150(50) & 51.1(3.3) & 0.9(6) & 319(44) & 0.09(32) & 74(156) \\
HD 153950 & 1 & 5.77 & 4.5 & 499.4(3.6) & 2.7(1) & 251(9) & 
0.34(2) & 308(2) & & & & & \\
          & 2 & 8.84 & 6.2 & 478.6(8.9) & 3.1(2) & 218(32) & 0.16(6) & 
303(31) & 205(67) & 0.4(3) & 206(56) & 0.10(31) & 188(29) \\
HD 154672 & 1 & 3.83 & 4.4 & 163.4(1) & 5.0(2) & 124(2) & 0.629(8) & 
265(1) & & & & & \\
          & 2 & 167.55 & 29.7 & 165.06(87) & 5.7(5) & 329(60) & 0.07(12) 
& 270(35) & 20.4(5) & 1.9(1.5) & 66(58) & 0.09(29) & 94(53) \\
HD 156279 & 1 & 8.18 & 9.7 & 131.1(5) & 9.7(4) & 184(2) & 0.71(2) & 
264(2) & & & & & \\
          & 2 & 1284.06 & 74.7 & 144(9) & 14(8) & 246(50) & 0.1(2) & 
209(41) & 151(8855) & 11(7) & 148(54) & 0.2(3) & 102(50) \\
HD 156846 & 1 & 5.18 & 25.8 & 359.3(1) & 10.9(3) & 205.3(4) & 0.846(2) & 
52.3(5) & & & & & \\
          & 2 & 1615.61 & 163.2 & 344.8(16.4) & 18(8) & 279(50) & 
0.00(29) & 322(27) & 175.7(1.7) & 11(3) & 36(60) & 0.12(5) & 332(51) \\
HD 171028 & 1 & 5.29 & 2.5 & 550(3) & 1.95(8) & 146(3) & 0.593(8) & 
304(1) & & & & & \\
          & 2 & 578.90 & 35.0 & 546.97 & 2.1(3) & 288(1) & 0.03 & 235 & 
546.90 & 2.1(3) & 288(1) & 0.03 & 55 \\
HD 171238 & 1 & 7.20 & 12.9 & 1466(33) & 2.8(2) & 250(22) & 0.26(4) &
75.7(9.7) & & & & & \\
          & 2 & 6.81 & 12.9 & 1517(105) & 3.2(2) & 288(36) & 0.20(6) & 
53(30) & 122.7(2.7) & 0.2(3) & 332(6) & 0.04(28) & 115(28) \\
HD 175167 & 1 & 2.70 & 5.3 & 1290(12) & 7.8(1.5) & 246(11) & 0.54(7) &
343(7) & & & & & \\
          & 2 & 6.95 & 7.9 & 1386(23) & 7.9(6) & 31(7) & 0.10 & 204 & 
302(3) & 2.0(3) & 0(14) & 0.16 & 126 \\
HD 175541 & 1 & 7.15 & 5.1 & 297.3(1.3) & 0.58(6) & 90(20) & 0.31(10) &
179(19) & & & & & \\
          & 2 & 6.72 & 4.3 & 295.0(1.7) & 0.6(1) & 79(38) & 0.09(14) & 
181(59) & 1180(70) & 0.4(9) & 124(32) & 0.13(23) & 204(38) \\
HD 187085 & 1 & 8.50 & 5.9 & 1032(11) & 0.87(8) & 24(38) & 0.11(7) &
120(37) & & & & & \\
          & 2 & 17.49 & 8.0 & 1031(14) & 0.9(8) & 347(62) & 0.07(8) & 
157(50) & 26.4(1.1) & 0.07(13) & 243(34) & 0.0(3) & 243(52) \\
HD 190228 & 1 & 0.78 & 7.4 & 1136(10) & 5.9(3) & 171(9) & 0.53(3) & 
101(2) & & & & & \\
          & 2 & 1.11 & 8.4 & 1108(26) & 7(5) & 123(42) & 0.17(26) & 
208(59) & 1110(19) & 7(5) & 208(59) & 0.20(17) & 346(54) \\
HD 192310 & 1 & 12.51 & 3.3 & 74.4(1) & 0.042(6) & 249(33) & 0.34(12) &
22(20) & & & & & \\
          & 2 & 5.99 & 1.9 & 74.4(2) & 0.05(1) & 51(41) & 0.04(14) & 
9(55) & 629(64) & 0.05(13) & 285(28) & 0.03(25) & 25(50) \\
HD 196885 & 1 & 5.11 & 19.5 & 1277(13) & 2.1(2) & 162(12) & 0.32(5) &
96(12) & & & & & \\
          & 2 & 77.38 & 78.3 & 1343(46) & 4(3) & 216(82) & 0.19(13) & 
69(24) & 391(5) & 3(8) & 250(30) & 0.2(2) & 128(50) \\
HD 204941 & 1 & 4.26 & 1.3 & 1595(67) & 0.26(4) & 131(43) & 0.14(9) &
265(29) & & & & & \\
          & 2 & 3.76 & 1.1 & 1696(119) & 0.23(2) & 55(23) & 0.07(9) & 
357(52) & 8.31(1) & 0.01(1) & 270(58) & 0.0(3) & 0(34) \\
HD 210277 & 1 & 2.04 & 6.8 & 442.16(35) & 1.29(5) & 141(2) & 
0.473(12) & 118(2) & & & & & \\
          & 2 & 6.92 & 8.6 & 443(40) & 1.2(5) & 358(35) & 0.04(44) & 
91(54) & 443.2(7) & 2.4(5) & 141(53) & 0.20(9) & 125(27) \\
HD 213240 & 1 & 4.47 & 10.9 & 872.74(96) & 4.4(2) & 161(1) & 
0.428(9) & 204.4(1.3) & & & & & \\
          & 2 & 12.03 & 11.7 & 870(4) & 4.6(2) & 336(64) & 0.06(3) & 
252(15) & 870(2) & 8.0(2) & 131(47) & 0.20(2) & 257(33) \\
HD 216437 & 1 & 8.28 & 5.8 & 1354(5) & 2.1(1) & 63(4) &
0.35(2) & 63(4) & & & & & \\
          & 2 & 10.32 & 6.1 & 1342(56) & 3(3) & 95(47) & 0.2(2) & 
1343(62) & 312(68) & 3(3) & 38(36) & 0.1(3) & 148(50) \\
HD 216770 & 1 & 3.83 & 8.6 & 118.4(9) & 0.64(7) & 216(27) &
0.38(11) & 280(20) & & & & & \\
          & 2 & 2.37 & 4.4 & 116.7(1.7) & 0.7(3) & 37(73) & 0.1(3) & 
69(27) & 41.1(3) & 0.3(2) & 237(51) & 0.0(3) & 131(62) \\
HD 217786 & 1 & 1.77 & 2.7 & 1314.7(3.4) & 13(1) & 137(2) & 
0.385(42) & 101.2(1.7) & & & & & \\
          & 2 & 2.68 & 3.0 & 1295 & 11(10) & 279(1) & 0.00(8) & 
315(62) & 631(11) & 1.0(7) & 59(45) & 0.11(8) & 298(31) \\
HD 218566 & 1 & 8.41 & 3.5 & 225.7(4) & 0.21(2) & 11(17) & 
0.29(7) & 36(18) & & & & & \\
          & 2 & 6.40 & 2.9 & 224.9(7) & 0.20(2) & 313(37) & 0.05(14) & 
87(25) & 1311(46) & 0.2(4) & 229(41) & 0.04(26) & 58(33) \\
HD 222582 & 1 & 1.80 & 3.7 & 572.3(7) & 7.6(4) & 62(2) & 0.73(2) & 
319(1) & & & & & \\
          & 2 & 109.44 & 28.3 & 586 & 9(7) & 31(42) & 0.05(20) & 135(31) 
& 573(3) & 14(5) & 111(33) & 0.2(2) & 242(25) \\
HD 240237 & 1 & 33.61 & 35.5 & 747(16) & 5.2(9) & 190(19) & 0.40(16) &
104(25) & & & & & \\
          & 2 & 25.03 & 29.5 & 753(36) & 5(3) & 185(44) & 0.0(2) & 
128(45) & 22.49(6) & 1(2) & 315(37) & 0.2(3) & 199(61) \\
HIP 2247 & 1 & 10.14 & 4.5 & 655.6(6) & 5.1(3) & 156(1) & 0.543(5) & 
112(2) & & & & & \\
         & 2 & 155.09 & 15.6 & 632(10) & 6.2(3) & 108(47) & 0.16(10) & 
130(42) & 75.7(1.2) & 1(1) & 343(28) & 0.20(21) & 344(42) \\
HIP 5158 & 1 & 9.90 & 10.0 & 352.6(7) & 1.55(2) & 8(8) & 0.537(fixed) & 
253(3) & & & & & \\
         & 2 & 23.04 & 6.2 & 385.7(7.9) & 2(3) & 156(61) & 0.18(17) & 
209(63) & 401(10) & 2(3) & 172(42) & 0.20(14) & 26(63) \\
HIP 57050 & 1 & 13.08 & 9.4 & 41.40(2) & 0.30(4) & 321(18) & 0.31(fixed) &
238(12) & & & & & \\
          & 2 & 12.95 & 8.6 & 41.40(2) & 0.34(5) & 318(36) & 0.1(1) & 
244(4) & 28.52(6) & 0.1(1) & 19(53) & 0.04(34) & 71(47) \\
iota Dra & 1 & 9.49 & 14.0 & 511.15(8) & 12.7(3) & 128.5(2) & 0.711(4) & 
91.9(7) & & & & \\
         & 2 & 525.35 & 99.7 & 510 & 14.8(5) & 160(61) & 0.00(6) & 
65(56) & 267 & 6(5) & 243(38) & 0.2(2) & 163(23) \\
GJ 649 & 1 & 11.03 & 4.4 & 602(8) & 0.35(6) & 195(24) & 0.32(12) &
7(27) & & & & & \\
       & 2 & 7.21 & 4.1 & 599(5) & 0.31(9) & 153(61) & 0.07(12) & 36(64) 
& 22.36(1) & 0.05(8) & 2(47) & 0.2(2) & 346(46) \\
GJ 676A & 1 & 4.03 & 3.7 & 1057(3) & 4.9(3) & 211(2) & 0.326(9) &
85.7(1.4) & & & & & \\
        & 2 & 6.80 & 4.6 & 1047(24) & 5(1) & 223(53) & 0.20(4) & 26(34) 
& 1126(270) & 4(1) & 185(41) & 0.2(1) & 203(64) \\
14 Her & 1 & 5.54 & 13.5 & 1754.7(4.3) & 5.2(3) & 328(2) & 0.388(9) &
23.2(1.6) & & & & & \\
         & 2 & 24.92 & 17.1 & 1740(696) & 6(8) & 290(42) & 0.19(17) & 
115(37) & 1755(917) & 5(7) & 2(45) & 0.19(14) & 278(37) \\
42 Dra & 1 & 51.21 & 26.8 & 480.2(3.2) & 3.7(3) & 288(6) & 0.51(6) &
210(7) & & & & & \\
       & 2 & 40.68 & 22.0 & 486(9) & 4.1(5) & 264(20) & 0.14(10) & 
236(55) & 69.9(6) & 0.5(9) & 138(62) & 0.11(23) & 35(51) \\
70 Vir & 1 & 1.63 & 7.7 & 116.686(4) & 7.4(2) & 339.9(7) & 0.399(3) &
358.9(4) & & & & & \\
       & 2 & 39.03 & 32.9 & 116.706(8) & 8.3(4) & 224(54) & 0.1995(82) & 
355(75) & 116.7(5.6) & 1.7(6) & 79(39) & 0.1(3) & 309(36) \\
\enddata 
\tablenotetext{a}{Single eccentric planet.} 
\tablenotetext{b}{Two planets, $e<0.2$.} 
\end{deluxetable} 

\begin{deluxetable}{llllllllllllllll}
\rotate
\tabletypesize{\scriptsize}
\tablecolumns{16}
\tablewidth{0pt}
\tablecaption{Two-Planet Fits }
\tablehead{
\colhead{Star} & \colhead{Mass} & \colhead{$\chi^{2}_{\nu}$} & 
\colhead{RMS} & \colhead{$P_1$} & \colhead{$m_1$ sin $i$} & 
\colhead{$a_1$} & \colhead{$M_1$} & \colhead{$e_1$} & 
\colhead{$\omega_1$} & \colhead{$P_2$} & \colhead{$m_2$ sin $i$} & 
\colhead{$a_2$} & \colhead{$M_2$} & \colhead{$e_2$} & 
\colhead{$\omega_2$} \\
\colhead{} & \colhead{\Msun} & \colhead{} & \colhead{(\ms)} & 
\colhead{(days)} & \colhead{(\Mjup)} & \colhead{(AU)} & 
\colhead{(degrees)} & \colhead{} & \colhead{degrees} & \colhead {(days)} 
& \colhead{(\Mjup)} & \colhead{(AU)} & \colhead{(degrees)} & \colhead{} 
& \colhead{(degrees)} \\
}
\startdata
\label{promising}
HD 3651 & 0.882 & 5.07 & 6.7 & 62.22(5) & 0.17(3) & 0.295(2) & 8(44) & 
0.06(20) & 18(43) & 31.08(2) & 0.09(3) & 0.186(2) & 318(42) & 0.04(20) & 
55(66) \\
HD 7449 & 1.05 & 81.96 & 5.7 & 1693(39) & 1.2(2) & 2.83(9) & 220(65) & 
0.13(11) & 323(64) & 615(19) & 0.4(2) & 1.44(5) & 354(32) & 0.0(2) & 
0(46) \\
HD 52265 & 1.17 & 2.19 & 11.2 & 119.1(4) & 1.05(4) & 0.499(5) & 253(57) 
& 0.0(1) & 359(54) & 59.9(2) & 0.35(09) & 0.316(3) & 303(75) & 
0.05(10) & 358(35) \\
HD 65216 & 0.92 & 1.96 & 7.2 & 572.4(2.1) & 1.26(4) & 1.30(3) & 89(51) & 
0.00(2) & 0(41) & 152.6(6) & 0.17(3) & 0.54(1) & 237(48) & 0.02(10) & 
0(54) \\
HD 85390 & 0.76 & 4.34 & 1.4 & 822(12) & 0.14(1) & 1.57(5) & 343(57) & 
0.00(8) & 0(49) & 3700(840) & 0.20(2) & 4.23(9) & 156(41) & 0.00(7) & 
0(49) \\
HD 89744 & 1.558 & 62.54 & 73.2 & 257.8(4) & 8.3(6) & 0.92(1) & 171(81) 
& 0.00(1) & 0(87) & 85.2(1) & 3.2(3) & 0.440(5) & 71(18) & 0.00(5) & 
0(86) \\
HD 92788 & 1.078 & 5.70 & 11.2 & 326(1) & 3.6(2) & 0.95(1) & 348(47) & 
0.00(11) & 6(46) & 162(3) & 0.9(3) & 0.60(1) & 66(39) & 0.04(21) & 0(20) 
\\
HD 117618 & 1.069 & 5.74 & 4.6 & 25.807(6) & 0.21(1) & 0.175(2) & 
217(35) & 0.00(8) & 0(48) & 318(2) & 0.2(1) & 0.93(1) & 304(50) & 
0.00(26) & 0(29) \\
GJ 649 & 0.54 & 4.19 & 3.2 & 601(6) & 0.33(5) & 1.14(5) & 211(49) & 
0.2(1) & 332(52) & 4.4762(4) & 0.030(8) & 0.043(1) & 334(68) & 0.20(15) 
& 334(47) \\
\enddata 
\end{deluxetable} 

\begin{deluxetable}{lrr}
\tabletypesize{\scriptsize}
\tablecolumns{3}
\tablewidth{0pt}
\tablecaption{AAT Radial Velocities for HD 2039}
\tablehead{
\colhead{JD-2400000} & \colhead{Velocity (\ms)} & \colhead{Uncertainty
(\ms)}}
\startdata
\label{2039vels}
51118.05806  &     29.4  &    4.2  \\
51118.96097  &      8.6  &    7.8  \\
51119.94453  &      5.7  &    5.2  \\
51121.03846  &     17.5  &    7.0  \\
51211.95142  &      5.6  &    8.8  \\
51212.92337  &      8.5  &    5.6  \\
51213.97494  &     26.2  &    7.2  \\
51214.91707  &      0.0  &    5.1  \\
51386.32274  &    -26.6  &    7.6  \\
51387.29810  &     -4.7  &    5.4  \\
51411.22931  &     15.5  &    8.0  \\
51414.25848  &    -16.2  &    4.7  \\
51473.08831  &    -43.5  &    4.5  \\
51525.92865  &    -53.8  &    8.0  \\
51527.92257  &    -46.7  &    5.8  \\
51745.27018  &    -68.7  &    9.9  \\
51828.07030  &    -23.6  &    6.6  \\
51828.99403  &    -34.2  &    5.9  \\
51829.97574  &    -28.3  &    6.6  \\
51856.07023  &    -26.5  &   10.3  \\
51919.94344  &     -4.8  &    7.1  \\
51920.96715  &     -8.3  &    6.9  \\
52093.29473  &    161.2  &    7.0  \\
52127.23412  &    104.9  &    7.2  \\
52151.22296  &     61.7  &    4.4  \\
52152.08603  &     68.2  &    3.9  \\
52154.21237  &     78.9  &    6.0  \\
52187.09574  &     62.2  &    4.4  \\
52188.03000  &     59.2  &    3.6  \\
52189.15021  &     55.5  &    5.3  \\
52190.09323  &     44.7  &    3.4  \\
52422.32809  &    -14.0  &    6.2  \\
52425.33222  &    -15.8  &    3.6  \\
52455.28482  &    -10.2  &    2.7  \\
52511.10451  &     -0.5  &    6.6  \\
52599.01528  &    -13.1  &    6.8  \\
53007.02913  &     -6.3  &    2.7  \\
53045.91967  &     -2.3  &    6.3  \\
53245.25642  &    112.7  &    3.6  \\
53579.22474  &      0.5  &    2.9  \\
54013.11160  &    -24.1  &    1.9  \\
54369.10873  &     79.4  &    2.3  \\
55102.09594  &    -43.0  &    3.6  \\
55430.23438  &    131.6  &    4.2  \\
55463.19594  &     95.8  &    5.8  \\
\enddata
\end{deluxetable}


\begin{deluxetable}{lrr}
\tabletypesize{\scriptsize}
\tablecolumns{3}
\tablewidth{0pt}
\tablecaption{AAT Radial Velocities for HD 20782}
\tablehead{
\colhead{JD-2400000} & \colhead{Velocity (\ms)} & \colhead{Uncertainty
(\ms)}}
\startdata
\label{20782vels}
51035.31946  &     21.7  &    2.3  \\
51236.93065  &     -6.7  &    3.3  \\
51527.01731  &      7.1  &    3.4  \\
51630.88241  &     29.5  &    2.7  \\
51768.30885  &     -6.8  &    2.6  \\
51828.11066  &     -7.8  &    3.0  \\
51829.27449  &     -6.8  &    3.8  \\
51829.99625  &    -26.7  &    8.7  \\
51856.13530  &    -10.5  &    3.5  \\
51919.00660  &     -3.8  &    2.9  \\
51919.99630  &     -1.8  &    2.9  \\
51983.89009  &      4.0  &    3.3  \\
52092.30437  &     17.7  &    2.4  \\
52127.26814  &     17.5  &    2.8  \\
52152.16308  &     23.0  &    2.5  \\
52187.15965  &     22.7  &    2.5  \\
52511.20613  &     -1.4  &    2.3  \\
52592.04802  &     17.3  &    2.3  \\
52654.96031  &     15.2  &    2.4  \\
52859.30540  &   -202.7  &    1.9  \\
52946.13848  &    -18.3  &    2.1  \\
52947.12256  &    -14.4  &    1.8  \\
53004.00130  &     -0.5  &    1.9  \\
53044.02367  &      0.5  &    2.2  \\
53045.96101  &     -0.6  &    1.9  \\
53217.28806  &      8.9  &    1.7  \\
53282.22016  &     20.5  &    1.9  \\
53398.96943  &     20.5  &    1.4  \\
53403.96080  &     28.5  &    2.6  \\
53576.30682  &     -9.4  &    1.6  \\
53632.28115  &     -7.9  &    1.6  \\
53665.18657  &      6.2  &    1.7  \\
54013.21634  &     31.1  &    1.5  \\
54040.13193  &     22.0  &    2.0  \\
54153.97021  &    -11.6  &    2.1  \\
54375.24665  &     13.2  &    1.7  \\
54544.89156  &     10.0  &    2.1  \\
54776.10133  &     -7.7  &    1.9  \\
54843.02061  &      0.1  &    1.6  \\
54899.92412  &     -0.8  &    2.1  \\
55107.24720  &     16.5  &    2.8  \\
55170.05454  &     17.1  &    2.4  \\
55204.97999  &     29.0  &    1.9  \\
55253.91182  &    -78.3  &    2.3  \\
55399.32278  &     -8.4  &    1.9  \\
55426.31464  &     -7.1  &    1.7  \\
55461.23926  &    -15.1  &    3.0  \\
55519.13337  &      8.2  &    2.0  \\
55844.13584  &   -144.3  &    6.6  \\
55845.17962  &   -187.0  &    2.3  \\
55846.13670  &   -156.4  &    2.3  \\
55964.93111  &      7.5  &    2.9  \\
\enddata
\end{deluxetable}


\begin{deluxetable}{lrr}
\tabletypesize{\scriptsize}
\tablecolumns{3}
\tablewidth{0pt}
\tablecaption{AAT Radial Velocities for HD 23127}
\tablehead{
\colhead{JD-2400000} & \colhead{Velocity (\ms)} & \colhead{Uncertainty
(\ms)}}
\startdata
\label{23127vels}
51118.09281  &    -20.9  &    5.2  \\
51119.17693  &     -7.8  &    6.0  \\
51120.26302  &    -65.5  &    8.8  \\
51121.12041  &    -40.5  &    4.8  \\
51157.11008  &    -26.3  &    4.2  \\
51211.97892  &    -65.7  &    4.9  \\
51212.95830  &    -37.5  &    4.0  \\
51213.99302  &    -31.5  &    3.9  \\
51214.94428  &    -47.4  &    3.3  \\
51473.24554  &     21.5  &    5.9  \\
51920.00758  &     -5.7  &    6.8  \\
51983.88167  &      9.1  &    6.2  \\
52092.31518  &     -0.4  &    4.7  \\
52127.28906  &     16.5  &    8.4  \\
52128.31303  &     17.8  &    8.9  \\
52151.30892  &     15.4  &    3.4  \\
52152.19974  &      0.0  &    3.9  \\
52188.15301  &      5.3  &    3.1  \\
52189.16917  &     -3.3  &    4.5  \\
52477.33581  &    -47.6  &   11.2  \\
52595.09050  &     -5.2  &    5.2  \\
52655.03366  &     -3.7  &    5.0  \\
52947.13705  &     16.2  &    2.0  \\
53004.02539  &     20.3  &    1.6  \\
53045.99621  &     11.0  &    2.3  \\
53217.30387  &      1.3  &    2.3  \\
53281.22070  &     -1.4  &    2.0  \\
53577.31797  &    -34.3  &    2.5  \\
53628.28608  &    -38.7  &    2.6  \\
53632.25573  &    -43.6  &    3.8  \\
53669.20119  &    -39.2  &    2.2  \\
53944.33250  &      2.7  &    1.9  \\
54010.19795  &      6.2  &    1.8  \\
54037.14466  &      4.0  &    2.1  \\
54374.24189  &     11.2  &    1.7  \\
54432.04206  &     31.4  &    4.3  \\
54552.89944  &     -3.1  &    2.8  \\
54905.91119  &    -37.7  &    2.8  \\
55105.20438  &    -16.3  &    2.9  \\
55172.05265  &     13.0  &    2.4  \\
55430.30950  &     21.1  &    2.9  \\
55522.11944  &     32.9  &    3.0  \\
55524.10270  &     18.4  &    8.9  \\
55967.93302  &     -4.9  &    3.4  \\
\enddata
\end{deluxetable}


\begin{deluxetable}{lrr}
\tabletypesize{\scriptsize}
\tablecolumns{3}
\tablewidth{0pt}
\tablecaption{AAT Radial Velocities for HD 38283}
\tablehead{
\colhead{JD-2400000} & \colhead{Velocity (\ms)} & \colhead{Uncertainty
(\ms)}}
\startdata
\label{38283vels}
50829.98715  &     -4.1  &    1.6  \\
50831.11229  &     -0.3  &    1.6  \\
51157.14186  &    -22.1  &    2.3  \\
51213.00749  &      3.1  &    2.6  \\
51526.07541  &    -16.5  &    2.6  \\
51530.13214  &     -7.4  &    3.0  \\
51683.84877  &      6.7  &    2.4  \\
51921.13329  &      6.0  &    1.9  \\
52188.26174  &     -4.4  &    2.4  \\
52594.17205  &     -8.1  &    2.2  \\
52654.09309  &      6.7  &    2.3  \\
52751.89940  &      4.2  &    2.1  \\
53004.05785  &     -6.7  &    1.6  \\
53042.04257  &      1.4  &    2.1  \\
53043.01088  &     -3.5  &    2.1  \\
53044.05527  &     -1.0  &    2.3  \\
53047.04258  &      5.5  &    1.9  \\
53048.08703  &      0.8  &    1.9  \\
53214.31960  &     -2.8  &    1.9  \\
53283.27433  &     -6.4  &    2.5  \\
53399.05143  &      4.5  &    1.6  \\
53483.85275  &     10.2  &    1.8  \\
53484.86059  &      4.2  &    1.7  \\
53486.87865  &      4.1  &    1.8  \\
53487.88987  &      3.9  &    1.7  \\
53842.85825  &      6.8  &    1.3  \\
54011.27866  &     -4.6  &    1.8  \\
54018.25770  &    -10.8  &    2.2  \\
54037.19283  &     -6.8  &    2.7  \\
54038.22781  &    -18.4  &    2.3  \\
54040.19869  &    -16.3  &    2.6  \\
54118.99517  &      0.1  &    1.4  \\
54221.85332  &      5.8  &    1.4  \\
54371.28261  &     -1.9  &    1.6  \\
54432.16897  &    -11.7  &    2.7  \\
54545.96278  &      4.5  &    1.6  \\
54777.16327  &    -18.1  &    2.4  \\
54780.22622  &     -7.9  &    1.6  \\
54899.97099  &     -1.3  &    2.4  \\
55105.27655  &    -10.2  &    2.3  \\
55171.10597  &     -5.7  &    1.7  \\
55201.13167  &     -7.4  &    2.1  \\
55205.04258  &      3.0  &    1.7  \\
55252.94918  &     10.7  &    1.7  \\
55309.87992  &     11.9  &    1.4  \\
55312.85776  &     15.3  &    1.9  \\
55315.85591  &      8.7  &    1.5  \\
55376.33979  &     -0.6  &    1.6  \\
55377.34088  &      1.8  &    1.9  \\
55398.31564  &     -8.5  &    1.3  \\
55401.33166  &     -9.2  &    1.6  \\
55457.28957  &     -4.9  &    2.0  \\
55462.26265  &     -5.7  &    2.2  \\
55519.20219  &     -2.7  &    1.9  \\
55524.14501  &      0.4  &    1.7  \\
55603.01744  &      8.8  &    1.9  \\
55665.89491  &     10.8  &    2.3  \\
55691.88048  &     23.2  &    1.7  \\
55874.20082  &    -12.4  &    2.1  \\
55898.15704  &     -2.5  &    2.3  \\
55963.02880  &     15.8  &    2.0  \\
\enddata
\end{deluxetable}


\begin{deluxetable}{lrr}
\tabletypesize{\scriptsize}
\tablecolumns{3}
\tablewidth{0pt}
\tablecaption{AAT Radial Velocities for HD 39091}
\tablehead{
\colhead{JD-2400000} & \colhead{Velocity (\ms)} & \colhead{Uncertainty
(\ms)}}
\startdata
\label{39091vels}
50829.99300  &      0.3  &    2.2  \\
51119.25037  &    -29.4  &    4.6  \\
51236.03289  &    -38.2  &    2.7  \\
51411.32492  &    -41.8  &    2.8  \\
51473.26697  &    -39.2  &    2.2  \\
51526.08042  &    -51.0  &    2.2  \\
51527.08206  &    -47.4  &    2.0  \\
51530.12796  &    -45.8  &    2.2  \\
51629.91162  &    -50.3  &    2.5  \\
51683.84224  &    -56.4  &    2.3  \\
51828.18751  &    -24.3  &    2.1  \\
51919.09891  &      2.8  &    3.3  \\
51921.13833  &     -0.4  &    2.3  \\
51983.91910  &     34.0  &    2.4  \\
52060.83961  &    178.8  &    2.1  \\
52092.33661  &    252.8  &    2.2  \\
52093.35148  &    253.3  &    2.0  \\
52127.32781  &    327.6  &    2.7  \\
52128.33566  &    327.9  &    2.0  \\
52130.33830  &    330.0  &    3.1  \\
52151.29169  &    349.2  &    2.2  \\
52154.30426  &    338.8  &    5.0  \\
52187.19587  &    325.7  &    1.8  \\
52188.23586  &    326.7  &    1.9  \\
52189.22191  &    320.9  &    1.7  \\
52190.14484  &    320.9  &    1.9  \\
52387.87064  &    140.3  &    1.7  \\
52510.30664  &     83.2  &    2.1  \\
52592.12558  &     59.7  &    1.5  \\
52599.15463  &     59.4  &    5.7  \\
52654.09858  &     56.8  &    2.3  \\
52708.98434  &     42.3  &   11.4  \\
52751.91773  &     29.8  &    2.1  \\
52944.22372  &     -1.5  &    1.9  \\
53004.07471  &      6.9  &    1.9  \\
53042.07800  &     -0.5  &    2.0  \\
53043.01734  &     -3.6  &    2.2  \\
53047.04936  &      0.0  &    2.1  \\
53048.09782  &     -6.3  &    1.7  \\
53245.31090  &    -29.2  &    2.5  \\
53402.03480  &    -21.7  &    0.9  \\
53669.24365  &    -44.6  &    1.0  \\
54012.24920  &      8.2  &    0.9  \\
54039.16869  &     16.9  &    1.1  \\
54224.85303  &    339.5  &    1.1  \\
54336.31438  &    246.9  &    1.8  \\
54337.29175  &    251.3  &    1.6  \\
54372.27038  &    208.0  &    1.8  \\
54425.22432  &    171.5  &    1.3  \\
54545.94257  &    104.4  &    1.1  \\
54841.06346  &     33.9  &    1.4  \\
54901.94078  &     19.7  &    2.0  \\
54905.99137  &     18.0  &    1.6  \\
54906.97424  &     18.5  &    1.3  \\
55106.23926  &     -4.7  &    1.9  \\
55170.23742  &     -6.3  &    1.3  \\
55202.05426  &     -1.5  &    1.9  \\
55252.96918  &    -19.1  &    1.4  \\
55521.20064  &    -32.6  &    1.6  \\
55585.99614  &    -23.1  &    1.7  \\
55664.85733  &    -31.0  &    1.5  \\
55846.25128  &    -35.0  &    1.5  \\
55898.10923  &    -34.1  &    1.3  \\
55899.13681  &    -35.7  &    1.4  \\
55962.00315  &    -23.3  &    1.3  \\
55965.03864  &    -24.6  &    1.4  \\
55966.97920  &    -22.9  &    2.3  \\
\enddata
\end{deluxetable}


\begin{deluxetable}{lrr}
\tabletypesize{\scriptsize}
\tablecolumns{3}
\tablewidth{0pt}
\tablecaption{AAT Radial Velocities for HD 102365}
\tablehead{
\colhead{JD-2400000} & \colhead{Velocity (\ms)} & \colhead{Uncertainty
(\ms)}}
\startdata
\label{102365vels}
50830.21201  &     -2.2  &    1.5  \\
50970.88818  &      0.2  &    1.2  \\
51213.22626  &     -7.9  &    1.4  \\
51236.20861  &      4.1  &    3.4  \\
51237.10884  &     -4.7  &    2.1  \\
51274.16691  &     -1.8  &    1.3  \\
51275.06855  &    -10.8  &    2.2  \\
51382.90174  &     -6.6  &    1.4  \\
51631.04270  &     -5.4  &    1.4  \\
51682.83774  &     -4.7  &    1.7  \\
51684.05028  &      1.0  &    1.5  \\
51717.85402  &     -3.6  &    1.5  \\
51743.86802  &     -2.0  &    1.5  \\
51919.23487  &      0.2  &    2.4  \\
51984.12770  &     -6.5  &    1.8  \\
52009.15814  &      0.0  &    1.4  \\
52060.92662  &     -1.6  &    1.5  \\
52127.86705  &     -3.8  &    2.0  \\
52388.04297  &     -7.1  &    1.5  \\
52420.97249  &      2.8  &    1.5  \\
52421.97640  &      1.6  &    1.5  \\
52422.90505  &     -0.8  &    1.6  \\
52423.97700  &     -0.5  &    1.8  \\
52424.97954  &     -1.0  &    1.1  \\
52455.88825  &     -7.4  &    1.5  \\
52654.27164  &      1.8  &    1.6  \\
52745.02160  &     -3.9  &    1.6  \\
52749.07992  &     -2.9  &    1.6  \\
52751.07418  &      1.0  &    1.6  \\
52783.96280  &      1.1  &    1.7  \\
52860.84417  &      3.7  &    1.6  \\
53005.25313  &      5.9  &    2.0  \\
53008.21509  &      1.7  &    1.6  \\
53041.28500  &      2.4  &    1.5  \\
53042.21355  &      0.2  &    1.5  \\
53048.25853  &      1.9  &    1.6  \\
53051.18853  &      5.3  &    1.5  \\
53214.87086  &      0.1  &    1.6  \\
53245.85116  &     -0.3  &    2.2  \\
53402.19479  &      0.2  &    0.8  \\
53482.94177  &      3.5  &    0.9  \\
53483.97344  &     -2.1  &    0.9  \\
53485.00857  &     -0.8  &    0.8  \\
53485.92867  &     -2.6  &    0.9  \\
53486.99104  &     -2.9  &    0.8  \\
53488.05937  &     -4.2  &    0.8  \\
53488.97513  &     -2.5  &    0.8  \\
53506.91912  &      2.5  &    0.9  \\
53509.01692  &      0.2  &    0.9  \\
53509.84592  &     -2.9  &    0.8  \\
53515.85480  &      4.0  &    0.8  \\
53516.84760  &      4.4  &    0.9  \\
53517.87645  &      2.0  &    0.8  \\
53518.93476  &      2.6  &    0.8  \\
53519.83473  &      2.5  &    0.9  \\
53520.97000  &      1.9  &    0.9  \\
53521.91817  &     -1.1  &    0.9  \\
53522.95210  &      0.4  &    0.9  \\
53568.84360  &     -3.6  &    0.8  \\
53569.86061  &     -2.1  &    0.9  \\
53570.88393  &     -4.4  &    0.9  \\
53571.89292  &     -5.1  &    0.9  \\
53572.86541  &     -1.6  &    0.9  \\
53573.84968  &     -4.2  &    0.8  \\
53575.85410  &     -1.2  &    0.8  \\
53576.85102  &     -1.8  &    0.8  \\
53577.84683  &     -1.5  &    0.8  \\
53578.84583  &     -0.2  &    0.8  \\
53700.24588  &      0.4  &    0.9  \\
53753.24814  &      0.1  &    1.1  \\
53840.11811  &     -7.6  &    1.2  \\
53841.00289  &     -1.4  &    0.8  \\
53844.01699  &     -1.1  &    0.9  \\
53937.87450  &      0.0  &    0.8  \\
54038.24696  &     -2.3  &    1.3  \\
54111.18437  &      4.2  &    0.9  \\
54112.19360  &      2.5  &    0.7  \\
54113.21647  &      3.4  &    0.9  \\
54114.22365  &      2.2  &    0.8  \\
54115.22923  &      1.0  &    1.1  \\
54119.23118  &      2.3  &    0.8  \\
54120.17724  &     -0.5  &    0.7  \\
54121.18308  &     -0.2  &    0.7  \\
54123.20774  &      2.9  &    0.7  \\
54126.14491  &     -1.8  &    0.7  \\
54127.15427  &      3.0  &    0.6  \\
54128.16856  &      0.7  &    0.9  \\
54129.17063  &     -1.2  &    0.6  \\
54130.16450  &      2.4  &    0.7  \\
54131.17111  &      1.7  &    0.7  \\
54132.17838  &      0.1  &    0.9  \\
54133.24020  &      2.0  &    1.0  \\
54134.20214  &      0.3  &    1.0  \\
54135.16844  &      2.3  &    0.8  \\
54136.18672  &      1.7  &    0.8  \\
54137.18488  &      2.0  &    0.7  \\
54138.16588  &      0.1  &    1.1  \\
54139.15652  &      1.9  &    0.8  \\
54140.15870  &      3.2  &    0.9  \\
54141.18313  &      3.9  &    0.9  \\
54142.17798  &      2.1  &    0.6  \\
54144.06348  &      1.3  &    0.7  \\
54145.14432  &      2.2  &    0.8  \\
54146.14936  &      0.7  &    0.7  \\
54147.18172  &     -1.0  &    0.7  \\
54148.21150  &     -1.0  &    0.8  \\
54149.15042  &     -2.1  &    0.8  \\
54150.14582  &     -1.9  &    0.7  \\
54151.19287  &     -1.9  &    0.7  \\
54152.20783  &     -0.2  &    0.9  \\
54153.15877  &     -3.6  &    0.9  \\
54154.10999  &     -3.0  &    0.6  \\
54155.07901  &     -1.9  &    0.7  \\
54156.05374  &     -2.5  &    1.1  \\
54222.04514  &     -1.5  &    1.7  \\
54223.06951  &      2.5  &    1.0  \\
54224.08922  &      0.6  &    0.9  \\
54225.03639  &     -1.9  &    0.8  \\
54226.00289  &     -2.1  &    1.3  \\
54252.95650  &     -0.1  &    0.9  \\
54254.90643  &      1.8  &    1.0  \\
54255.92803  &     -3.5  &    1.0  \\
54257.05236  &     -4.5  &    1.1  \\
54543.11240  &     -3.2  &    0.8  \\
54550.07875  &      0.1  &    1.4  \\
54551.04141  &     -2.4  &    1.0  \\
54553.06907  &     -6.7  &    1.1  \\
54841.22838  &      2.6  &    1.1  \\
54843.26009  &      0.2  &    1.1  \\
54897.14254  &      0.7  &    1.0  \\
54901.13378  &     -3.1  &    1.1  \\
54902.14792  &     -0.8  &    1.1  \\
54904.17743  &     -7.3  &    1.4  \\
54905.17801  &     -6.5  &    0.9  \\
54906.19747  &     -4.1  &    1.1  \\
54908.17853  &     -6.3  &    1.0  \\
55031.89041  &     -4.5  &    0.9  \\
55202.19583  &      6.7  &    1.2  \\
55204.23690  &      1.6  &    1.4  \\
55206.17590  &      0.3  &    1.0  \\
55231.14629  &      2.7  &    1.1  \\
55253.16083  &      2.9  &    1.0  \\
55310.04537  &     -2.6  &    1.2  \\
55312.05141  &     -2.7  &    1.0  \\
55313.06557  &      3.6  &    1.1  \\
55314.97472  &     -3.4  &    1.0  \\
55316.99125  &     -3.3  &    1.1  \\
55370.88444  &      1.1  &    1.2  \\
55371.88362  &     -6.5  &    1.0  \\
55374.93147  &      4.5  &    1.5  \\
55376.86882  &     -2.7  &    1.2  \\
55397.85721  &     -4.6  &    1.1  \\
55398.85407  &     -4.8  &    0.9  \\
55586.17916  &      8.4  &    1.1  \\
55603.28296  &     12.5  &    1.4  \\
55604.09249  &      8.6  &    0.9  \\
55664.07027  &      9.6  &    1.0  \\
55666.02599  &      1.2  &    1.0  \\
55692.02182  &     12.0  &    1.2  \\
55692.99384  &      8.4  &    1.1  \\
55750.86140  &      4.1  &    1.1  \\
55751.84824  &      9.1  &    2.2  \\
55753.83346  &      7.7  &    1.0  \\
55756.86016  &      5.2  &    1.6  \\
55785.87243  &     -0.7  &    1.3  \\
55787.84349  &     10.7  &    1.2  \\
55878.26267  &     -1.7  &    2.7  \\
55961.14964  &     11.2  &    1.0  \\
\enddata
\end{deluxetable}


\begin{deluxetable}{lrr}
\tabletypesize{\scriptsize}
\tablecolumns{3}
\tablewidth{0pt}
\tablecaption{AAT Radial Velocities for HD 108147}
\tablehead{
\colhead{JD-2400000} & \colhead{Velocity (\ms)} & \colhead{Uncertainty
(\ms)}}
\startdata
\label{108147vels}
50830.24279  &     26.4  &    2.6  \\
50915.04863  &     -7.8  &    4.0  \\
51213.25018  &     10.1  &    4.4  \\
51276.06356  &     21.0  &    4.3  \\
51382.88648  &    -43.0  &    2.5  \\
51631.05277  &    -28.9  &    2.6  \\
51682.96554  &     -6.1  &    2.7  \\
51718.00741  &    -10.1  &    2.8  \\
51856.26359  &     16.1  &    4.7  \\
51984.09097  &     34.3  &    4.0  \\
52009.12841  &     -7.8  &    3.0  \\
52010.19667  &    -25.6  &    2.9  \\
52061.02192  &     38.4  &    2.8  \\
52091.89265  &      0.2  &    2.8  \\
52126.88255  &    -19.0  &    9.5  \\
52127.87323  &      0.8  &    3.1  \\
52129.89531  &     -6.9  &    2.9  \\
52360.17918  &    -13.8  &    3.6  \\
52387.01113  &     -4.5  &    2.1  \\
52388.03800  &     21.6  &    2.1  \\
52389.03674  &      9.8  &    2.3  \\
52390.02488  &      0.7  &    2.3  \\
52420.97941  &     30.1  &    2.2  \\
52421.99109  &     17.2  &    2.3  \\
52423.06610  &      1.9  &    2.4  \\
52423.98213  &      9.0  &    2.5  \\
52424.98647  &     -5.5  &    2.6  \\
52452.95259  &     18.1  &    3.3  \\
52454.90924  &      0.2  &    2.7  \\
52455.90437  &      4.5  &    2.9  \\
52509.85414  &      5.2  &    2.2  \\
52510.85405  &     -5.2  &    2.4  \\
52598.25801  &     46.1  &    6.5  \\
52599.25428  &     40.1  &    7.4  \\
52654.25245  &    -16.0  &    3.0  \\
52655.17027  &    -12.7  &    3.1  \\
52710.11861  &    -21.0  &    4.2  \\
52710.93403  &    -16.3  &    4.1  \\
52712.05281  &    -21.7  &    3.8  \\
52745.08065  &    -34.3  &    2.9  \\
52746.05949  &    -12.0  &    3.1  \\
52748.02643  &     29.3  &    2.5  \\
52749.08757  &      6.0  &    2.7  \\
52750.03644  &    -12.7  &    2.7  \\
52751.08725  &    -11.5  &    2.6  \\
52752.04493  &    -13.0  &    2.6  \\
52783.95060  &     10.7  &    2.0  \\
52785.05259  &      1.1  &    2.6  \\
52785.97642  &    -17.7  &    2.6  \\
52857.86852  &     15.8  &    2.7  \\
52858.87366  &     -2.2  &    2.7  \\
52859.85266  &      3.6  &    3.2  \\
52860.85516  &      0.0  &    2.5  \\
53516.99066  &     -2.5  &    3.5  \\
54899.12290  &    -12.7  &    4.1  \\
55962.19009  &     -9.6  &    3.4  \\
55996.08993  &     45.3  &    3.3  \\
\enddata
\end{deluxetable}


\begin{deluxetable}{lrr}
\tabletypesize{\scriptsize}
\tablecolumns{3}
\tablewidth{0pt}
\tablecaption{AAT Radial Velocities for HD 117618}
\tablehead{
\colhead{JD-2400000} & \colhead{Velocity (\ms)} & \colhead{Uncertainty
(\ms)}}
\startdata
\label{117618vels}
50831.18597  &    -10.3  &    2.4  \\
50917.10104  &     10.1  &    3.6  \\
50970.94927  &     17.1  &    2.8  \\
51212.20608  &    -12.2  &    3.1  \\
51236.22669  &      2.8  &    6.7  \\
51274.24419  &      0.6  &    3.7  \\
51383.93108  &      2.7  &    2.5  \\
51386.85838  &      3.1  &    2.5  \\
51631.25935  &    -28.3  &    2.4  \\
51682.97674  &    -15.0  &    2.8  \\
51718.03450  &      4.3  &    2.9  \\
51920.26309  &      6.2  &    3.4  \\
51984.10352  &    -17.1  &    4.2  \\
52092.96337  &    -15.1  &    2.5  \\
52129.00532  &      5.0  &    4.2  \\
52387.04015  &      8.4  &    1.9  \\
52388.07932  &     11.9  &    2.2  \\
52422.00889  &     -0.3  &    2.0  \\
52452.97667  &    -10.6  &    1.9  \\
52455.92575  &     -4.3  &    2.1  \\
52509.87274  &    -13.7  &    1.9  \\
52510.87230  &     -6.2  &    2.0  \\
52710.17758  &     -8.6  &    1.8  \\
52710.96784  &     -4.1  &    2.0  \\
52712.07590  &    -10.6  &    1.8  \\
52745.14346  &     11.4  &    2.3  \\
52750.10349  &     14.2  &    1.8  \\
52752.08891  &     13.0  &    1.9  \\
52784.00059  &      2.6  &    3.1  \\
52785.06453  &     -3.7  &    1.7  \\
52785.98821  &     -6.8  &    1.8  \\
52857.88042  &     12.0  &    1.7  \\
53006.24282  &      7.1  &    1.8  \\
53007.24111  &      1.1  &    2.4  \\
53008.23805  &     12.6  &    1.6  \\
53041.23361  &      7.6  &    2.6  \\
53042.22925  &     -4.9  &    1.8  \\
53044.16694  &     -9.1  &    2.2  \\
53045.27837  &    -11.9  &    2.1  \\
53046.08722  &    -12.0  &    2.3  \\
53046.27982  &    -10.5  &    2.7  \\
53047.20181  &    -15.0  &    1.8  \\
53051.19446  &    -10.6  &    1.9  \\
53213.99332  &      9.6  &    1.5  \\
53214.89437  &      8.3  &    1.6  \\
53215.89127  &      9.0  &    1.9  \\
53216.92640  &      8.2  &    1.8  \\
53242.90299  &      3.0  &    1.7  \\
53244.94716  &      3.8  &    2.2  \\
53245.88103  &      5.1  &    1.8  \\
53399.20905  &     11.4  &    1.5  \\
53405.21458  &     -5.6  &    1.5  \\
53483.04532  &    -12.1  &    2.5  \\
53485.09022  &     -8.7  &    1.9  \\
53507.02859  &     -1.6  &    1.9  \\
53521.98706  &     13.9  &    1.7  \\
53568.94925  &     -9.3  &    1.7  \\
53576.90329  &     -1.1  &    1.5  \\
53943.90016  &     -5.1  &    1.3  \\
54144.17378  &      9.5  &    1.8  \\
54224.16387  &      2.6  &    1.8  \\
54254.02748  &     -1.5  &    1.6  \\
54545.13700  &    -25.7  &    1.5  \\
54897.21961  &     10.0  &    1.9  \\
54904.21623  &    -18.7  &    3.0  \\
55313.10430  &      1.2  &    1.6  \\
55376.93152  &      1.8  &    1.7  \\
55402.89503  &     -6.5  &    1.8  \\
55665.16541  &     23.3  &    1.7  \\
55964.26135  &     -1.9  &    1.6  \\
\enddata
\end{deluxetable}


\begin{deluxetable}{lrr}
\tabletypesize{\scriptsize}
\tablecolumns{3}
\tablewidth{0pt}
\tablecaption{AAT Radial Velocities for HD 142415}
\tablehead{
\colhead{JD-2400000} & \colhead{Velocity (\ms)} & \colhead{Uncertainty
(\ms)}}
\startdata
\label{142415vels}
52390.12777  &     29.8  &    1.8  \\
52422.09531  &     24.7  &    1.6  \\
52425.10054  &      4.3  &    2.0  \\
52453.01048  &     -4.2  &    2.3  \\
52476.98791  &      9.2  &    2.3  \\
52745.18623  &     14.5  &    2.6  \\
52751.17456  &     15.1  &    2.5  \\
52858.91808  &     12.6  &    2.4  \\
53486.12760  &    -97.2  &    2.4  \\
53509.16629  &     -9.1  &    2.0  \\
53520.12129  &      2.3  &    3.0  \\
53576.98888  &     33.9  &    2.5  \\
53844.17161  &    -37.3  &    1.6  \\
53944.98514  &     11.3  &    1.7  \\
54226.07632  &      0.0  &    1.6  \\
54256.02792  &    -40.8  &    2.9  \\
54373.88492  &     78.9  &    2.2  \\
54544.25079  &     44.1  &    1.8  \\
55020.87803  &    -21.1  &    2.3  \\
55043.95806  &    -32.6  &    1.6  \\
55054.91417  &    -34.8  &    1.5  \\
55076.98252  &    -58.2  &    2.2  \\
\enddata
\end{deluxetable}


\begin{deluxetable}{lrr}
\tabletypesize{\scriptsize}
\tablecolumns{3}
\tablewidth{0pt}
\tablecaption{AAT Radial Velocities for HD 187085}
\tablehead{
\colhead{JD-2400000} & \colhead{Velocity (\ms)} & \colhead{Uncertainty
(\ms)}}
\startdata
\label{187085vels}
51120.91699  &    -14.0  &    2.5  \\
51411.07528  &      1.8  &    3.4  \\
51683.16928  &     22.4  &    2.9  \\
51743.04943  &     15.8  &    3.0  \\
51767.00464  &     13.7  &    2.4  \\
51769.06523  &      4.7  &    2.2  \\
51770.11535  &     11.7  &    2.6  \\
51855.94773  &      7.1  &    4.3  \\
52061.21399  &     -2.2  &    2.6  \\
52092.05117  &    -24.1  &    2.9  \\
52128.02674  &    -18.8  &    2.5  \\
52151.01458  &    -10.6  &    3.2  \\
52189.92296  &     -8.4  &    2.0  \\
52360.28155  &     -6.1  &    2.5  \\
52387.21676  &     -9.1  &    1.9  \\
52388.23551  &    -10.7  &    2.1  \\
52389.26643  &    -12.9  &    2.3  \\
52422.21212  &     -3.1  &    2.2  \\
52456.09167  &     -6.0  &    2.5  \\
52750.25882  &     11.8  &    1.9  \\
52752.23100  &     13.6  &    2.0  \\
52784.20681  &     23.0  &    1.8  \\
52857.12186  &      3.5  &    1.9  \\
52861.01669  &     16.4  &    2.3  \\
52942.98422  &     17.3  &    2.2  \\
52946.92563  &     18.9  &    1.8  \\
53217.06633  &     -9.0  &    2.0  \\
53245.04289  &    -21.2  &    2.8  \\
53484.30018  &     -1.5  &    1.9  \\
53489.26010  &     -6.6  &    1.7  \\
53507.19655  &      2.2  &    1.4  \\
53510.21563  &     -2.6  &    1.8  \\
53517.26209  &     -2.5  &    2.2  \\
53520.27478  &     -4.6  &    2.2  \\
53569.08540  &      0.1  &    2.0  \\
53572.16686  &     -3.2  &    2.2  \\
53577.02722  &      6.2  &    1.6  \\
53627.96289  &     14.7  &    3.2  \\
53632.07459  &     10.7  &    1.7  \\
53665.95368  &     20.3  &    1.9  \\
53945.13814  &     11.3  &    1.4  \\
54008.96665  &      8.3  &    1.6  \\
54016.93647  &      4.8  &    1.7  \\
54225.29248  &    -19.5  &    2.0  \\
54254.14742  &     -8.7  &    1.8  \\
54338.14695  &    -15.1  &    1.7  \\
54371.96418  &     -4.0  &    1.6  \\
54544.28035  &    -11.0  &    1.9  \\
54779.95738  &     11.3  &    2.1  \\
55101.98352  &      8.3  &    1.8  \\
55105.98996  &     -8.0  &    2.4  \\
55109.00947  &      3.6  &    1.8  \\
55110.96110  &      3.8  &    2.3  \\
55111.95166  &      3.9  &    1.6  \\
55313.29333  &    -18.6  &    1.6  \\
55315.24843  &    -17.4  &    1.8  \\
55317.23004  &    -17.8  &    2.1  \\
55376.17401  &    -16.9  &    1.9  \\
55399.11015  &    -16.1  &    2.3  \\
55430.02150  &     -3.9  &    2.1  \\
55462.01469  &     -1.7  &    2.7  \\
55664.31796  &     12.3  &    1.7  \\
55755.04339  &     11.1  &    1.9  \\
55844.91980  &     33.1  &    2.4  \\
\enddata
\end{deluxetable}


\begin{deluxetable}{lrr}
\tabletypesize{\scriptsize}
\tablecolumns{3}
\tablewidth{0pt}
\tablecaption{AAT Radial Velocities for HD 213240}
\tablehead{
\colhead{JD-2400000} & \colhead{Velocity (\ms)} & \colhead{Uncertainty
(\ms)}}
\startdata
\label{213240vels}
51034.19977  &     45.8  &    2.0  \\
51119.03196  &     19.7  &    5.0  \\
51683.29093  &     23.0  &    2.1  \\
51745.21846  &     43.6  &    2.3  \\
51767.18008  &     45.3  &    1.8  \\
51768.20681  &     43.2  &    2.3  \\
51856.03358  &     44.1  &    3.1  \\
51856.90638  &     48.4  &    4.9  \\
52010.31187  &     23.8  &    2.1  \\
52062.32344  &     14.9  &    1.9  \\
52092.20773  &     15.4  &    2.1  \\
52093.21992  &      0.0  &    2.0  \\
52127.17837  &      6.5  &    2.3  \\
52151.12818  &     -6.9  &    1.9  \\
52186.94703  &    -23.2  &    1.5  \\
52188.04540  &    -20.9  &    2.1  \\
52189.03997  &    -22.8  &    1.1  \\
52189.96724  &    -25.8  &    1.5  \\
52388.30013  &   -133.9  &    1.8  \\
52389.30681  &   -137.5  &    3.2  \\
52421.33402  &    -99.2  &    1.7  \\
52423.30063  &    -95.0  &    1.8  \\
52425.31971  &    -91.5  &    1.6  \\
52477.19309  &    -24.4  &    2.1  \\
52511.01295  &     12.8  &    2.0  \\
52593.94991  &     31.5  &    3.4  \\
52861.16704  &     34.1  &    2.2  \\
53216.25264  &   -133.0  &    2.0  \\
53244.17584  &   -145.3  &    1.9  \\
53576.18990  &     45.9  &    1.6  \\
54013.02040  &    -73.8  &    1.2  \\
54256.22716  &      9.2  &    1.5  \\
54776.91931  &    -23.0  &    1.9  \\
55102.03566  &    -19.1  &    1.6  \\
55520.94864  &     36.3  &    1.9  \\
\enddata
\end{deluxetable}


\begin{deluxetable}{lrr}
\tabletypesize{\scriptsize}
\tablecolumns{3}
\tablewidth{0pt}
\tablecaption{AAT Radial Velocities for HD 216437}
\tablehead{
\colhead{JD-2400000} & \colhead{Velocity (\ms)} & \colhead{Uncertainty
(\ms)}}
\startdata
\label{216437vels}
50830.94196  &    -16.0  &    1.6  \\
51034.22506  &    -16.3  &    1.8  \\
51386.30509  &     10.0  &    2.2  \\
51472.95520  &     20.2  &    1.8  \\
51683.31464  &     49.4  &    1.9  \\
51684.32758  &     48.7  &    1.8  \\
51743.23428  &     61.1  &    2.9  \\
51767.20463  &     58.8  &    1.7  \\
51768.22482  &     54.0  &    1.9  \\
51828.04269  &     65.4  &    2.0  \\
51828.96337  &     56.3  &    2.0  \\
51829.95683  &     55.0  &    2.1  \\
51856.04780  &     48.6  &    3.7  \\
51919.92935  &     47.8  &    1.8  \\
51920.92549  &     50.4  &    2.1  \\
52061.28844  &     -2.5  &    2.0  \\
52092.22059  &     -5.6  &    1.9  \\
52127.19806  &     -3.5  &    2.2  \\
52154.10649  &    -17.0  &    2.0  \\
52188.08067  &    -31.4  &    1.4  \\
52387.31939  &     -6.7  &    1.7  \\
52388.30982  &     -8.7  &    1.0  \\
52389.29622  &    -12.9  &    3.1  \\
52390.31831  &    -13.2  &    2.1  \\
52422.30856  &    -15.1  &    1.7  \\
52425.32603  &    -11.6  &    1.6  \\
52456.28016  &    -10.9  &    2.0  \\
52477.20241  &     -9.0  &    2.2  \\
52511.03790  &     -5.5  &    1.9  \\
52594.93793  &     -1.8  &    1.5  \\
52861.20216  &     32.8  &    2.2  \\
52945.03284  &     36.5  &    2.0  \\
53006.97127  &     44.8  &    1.9  \\
53215.25326  &     56.1  &    1.4  \\
53244.19252  &     52.5  &    2.2  \\
53509.30457  &     -9.1  &    0.9  \\
53523.32674  &    -18.2  &    1.2  \\
53577.20037  &    -21.6  &    0.9  \\
53632.15021  &    -24.1  &    0.9  \\
53943.23990  &      0.3  &    0.7  \\
54012.06760  &      2.7  &    0.8  \\
54014.10754  &      3.2  &    1.1  \\
54255.20770  &     26.3  &    1.1  \\
54375.09160  &     39.1  &    0.9  \\
54427.02974  &     43.6  &    3.2  \\
54752.04756  &      0.1  &    0.8  \\
55106.11746  &    -25.4  &    1.3  \\
55171.99939  &    -12.1  &    1.3  \\
55523.92910  &     22.3  &    1.0  \\
55751.28024  &     45.5  &    2.0  \\
\enddata
\end{deluxetable}



\begin{thebibliography}{}

\bibitem[Anglada-Escud{\'e} et al.(2010)]{ang10} Anglada-Escud{\'e}, G., 
L{\'o}pez-Morales, M., \& Chambers, J.~E.\ 2010, \apj, 709, 168

\bibitem[Arriagada et al.(2010)]{arr10} Arriagada, P., Butler, R.~P., 
Minniti, D., et al.\ 2010, \apj, 711, 1229

\bibitem[Butler et al.(2006)]{butler06} Butler, R.~P., et al.\
2006, \apj, 646, 505

\bibitem[Chambers(1999)]{chambers99} Chambers, J.~E.\ 1999, \mnras, 304, 
793

\bibitem[Chambers et al.(1996)]{chambers96} Chambers, J.~E., Wetherill, 
G.~W., \& Boss, A.~P.\ 1996, Icarus, 119, 261

\bibitem[Cochran et al.(2007)]{cochran07} Cochran, W.~D., Endl, M., 
Wittenmyer, R.~A., \& Bean, J.~L.\ 2007, ApJ, 665, 1407

\bibitem[Cochran et al.(2004)]{37605} Cochran, W.~D., Endl, M., 
McArthur, B., et al.\ 2004, \apjl, 611, L133

\bibitem[Correia et al.(2008)]{corr08} Correia, A.~C.~M., 
Udry, S., Mayor, M., et al.\ 2008, \aap, 479, 271

\bibitem[Cumming et al.(2008)]{cumming08} Cumming, A., Butler, R.~P., 
Marcy, G.~W., Vogt, S.~S., Wright, J.~T., \& Fischer, D.~A.\ 2008, 
\pasp, 120, 531

\bibitem[Cresswell \& Nelson(2009)]{cress09} Cresswell, P., \& Nelson, 
R.~P.\ 2009, \aap, 493, 1141

\bibitem[da Silva et al.(2006)]{dasilva06} da Silva, R., Udry, 
S., Bouchy, F., et al.\ 2006, \aap, 446, 717

\bibitem[da Silva et al.(2007)]{dasilva07} da Silva, R., Udry, 
S., Bouchy, F., et al.\ 2007, \aap, 473, 323

\bibitem[D{\'{\i}}az et al.(2012)]{diaz12} D{\'{\i}}az, 
R.~F., Santerne, A., Sahlmann, J., et al.\ 2012, \aap, 538, A113

\bibitem[D{\"o}llinger et al.(2009)]{dol09} D{\"o}llinger, 
M.~P., Hatzes, A.~P., Pasquini, L., et al.\ 2009, \aap, 499, 935

\bibitem[Dumusque et al.(2012)]{alphacen} Dumusque, X., Pepe, F., Lovis, 
C., et al.\ 2012, \nat, 491, 207

\bibitem[Dumusque et al.(2011)]{dum11} Dumusque, X., 
Lovis, C., S{\'e}gransan, D., et al.\ 2011, \aap, 535, A55

\bibitem[Eggenberger et al.(2006)]{egg06} Eggenberger, A., 
Mayor, M., Naef, D., et al.\ 2006, \aap, 447, 1159

\bibitem[Endl et al.(2004)]{endl04} Endl, M., Hatzes, A.~P., Cochran, 
W.~D., et al.\ 2004, \apj, 611, 1121

\bibitem[Endl et al.(2006)]{endl06} Endl, M., Cochran, W.~D., 
Wittenmyer, R.~A., \& Hatzes, A.~P.\ 2006, \aj, 131, 3131

\bibitem[Fischer et al.(2009)]{fischer09} Fischer, D., 
Driscoll, P., Isaacson, H., et al.\ 2009, \apj, 703, 1545

\bibitem[Fischer et al.(2007)]{fischer07} Fischer, D.~A., Vogt, S.~S., 
Marcy, G.~W., et al.\ 2007, \apj, 669, 1336

\bibitem[Ford(2008)]{scheduling} Ford, E.~B.\ 2008, \aj, 135, 1008

\bibitem[Ford \& Rasio(2008)]{ford08} Ford, E.~B., \& 
Rasio, F.~A.\ 2008, \apj, 686, 621

\bibitem[Forveille et al.(2011)]{for11} Forveille, T., 
Bonfils, X., Lo Curto, G., et al.\ 2011, \aap, 526, A141

\bibitem[Frink et al.(2002)]{frink02} Frink, S., Mitchell, D.~S., 
Quirrenbach, A., et al.\ 2002, \apj, 576, 478

\bibitem[Funk et al.(2012)]{funk12} Funk, B., Schwarz, R., S{\"u}li, 
{\'A}., \& {\'E}rdi, B.\ 2012, \mnras, 423, 3074

\bibitem[Gettel et al.(2012)]{gettel12} Gettel, S., Wolszczan, A., 
Niedzielski, A., et al.\ 2012, \apj, 745, 28

\bibitem[Giuppone et al.(2012)]{g12} Giuppone, C.~A., 
Ben{\'{\i}}tez-Llambay, P., \& Beaug{\'e}, C.\ 2012, \mnras, 421, 356

\bibitem[Go{\'z}dziewski \& Konacki(2006)]{goz06} 
Go{\'z}dziewski, K., \& Konacki, M.\ 2006, \apj, 647, 573

\bibitem[Haghighipour et al.(2010)]{hag10} Haghighipour, N., Vogt, 
S.~S., Butler, R.~P., et al.\ 2010, \apj, 715, 271

\bibitem[Hollis et al.(2012)]{hollis12} Hollis, M.~D.~J., Balan, S.~T., 
Lever, G., \& Lahav, O.\ 2012, \mnras, 423, 2800

\bibitem[Horner \& Lykawka(2010)]{jonti1} Horner, J., \& 
Lykawka, P.~S.\ 2010, \mnras, 405, 49

\bibitem[Horner et al.(2012a)]{jonti2} Horner, J., Lykawka, P.~S., 
Bannister, M.~T., \& Francis, P.\ 2012a, \mnras, 422, 2145

\bibitem[Horner et al.(2012b)]{NNSer} Horner, J., Wittenmyer, 
R.~A., Hinse, T.~C., \& Tinney, C.~G.\ 2012b, \mnras, 425, 749 

\bibitem[Horner et al.(2012c)]{HWVir} Horner, J., Hinse, T.~C., 
Wittenmyer, R.~A., Marshall, J.~P., \& Tinney, C.~G.\ 2012c, 
\mnras, 427, 2812

\bibitem[Howard et al.(2010a)]{howard10a} Howard, A.~W., Johnson, J.~A., 
Marcy, G.~W., et al.\ 2010a, \apj, 721, 1467

\bibitem[Howard et al.(2010b)]{howard10} Howard, A.~W., et al.\ 2010b, 
Science, 330, 653

\bibitem[Howard et al.(2011)]{howard11} Howard, A.~W., Johnson, J.~A., 
Marcy, G.~W., et al.\ 2011, \apj, 730, 10

\bibitem[Jenkins et al.(2009)]{jenkins09} Jenkins, J.~S., Jones, 
H.~R.~A., Go{\'z}dziewski, K., et al.\ 2009, \mnras, 398, 911

\bibitem[Johnson et al.(2006)]{johnson06} Johnson, J.~A., Marcy, G.~W., 
Fischer, D.~A., et al.\ 2006, \apj, 647, 600

\bibitem[Johnson et al.(2007)]{johnson07} Johnson, J.~A., Fischer, 
D.~A., Marcy, G.~W., et al.\ 2007, \apj, 665, 785

\bibitem[Johnson et al.(2010)]{johnson10} Johnson, J.~A., Howard, 
A.~W., Marcy, G.~W., et al.\ 2010, \pasp, 122, 149 

\bibitem[Johnson et al.(2011)]{johnson11} Johnson, J.~A., Clanton, C., 
Howard, A.~W., et al.\ 2011, \apjs, 197, 26

\bibitem[Jones et al.(2002)]{jones02} Jones, H.~R.~A., Paul Butler, R., 
Marcy, G.~W., et al.\ 2002, \mnras, 337, 1170

\bibitem[Jones et al.(2006)]{jones06} Jones, H.~R.~A., Butler, R.~P., 
Tinney, C.~G., et al.\ 2006, \mnras, 369, 249

\bibitem[Laughlin \& Chambers(2002)]{laughlin02} Laughlin, G., \& 
Chambers, J.~E.\ 2002, \aj, 124, 592

\bibitem[Levison et al.(1997)]{levison97} Levison, H.~F., 
Shoemaker, E.~M., \& Shoemaker, C.~S.\ 1997, \nat, 385, 42

\bibitem[Lo Curto et al.(2010)]{locurto10} Lo Curto, G., 
Mayor, M., Benz, W., et al.\ 2010, \aap, 512, A48

\bibitem[L{\'o}pez-Morales et al.(2008)]{lopez08} L{\'o}pez-Morales, M., 
Butler, R.~P., Fischer, D.~A., et al.\ 2008, \aj, 136, 1901

\bibitem[Marcy et al.(2005)]{marcy05} Marcy, G.~W., Butler, R.~P., Vogt, 
S.~S., et al.\ 2005, \apj, 619, 570

\bibitem[Mayor et al.(2004)]{mayor04} Mayor, M., Udry, S., 
Naef, D., et al.\ 2004, \aap, 415, 391

\bibitem[Meschiari et al.(2009)]{mes09} Meschiari, S., Wolf, A.~S., 
Rivera, E., et al.\ 2009, \pasp, 121, 1016

\bibitem[Meschiari et al.(2011)]{mes11} Meschiari, S., Laughlin, G., 
Vogt, S.~S., et al.\ 2011, \apj, 727, 117

\bibitem[Mordasini et al.(2011)]{mordasini11} Mordasini, C., 
Mayor, M., Udry, S., et al.\ 2011, \aap, 526, A111

\bibitem[Moutou et al.(2009)]{moutou09} Moutou, C., Mayor, 
M., Lo Curto, G., et al.\ 2009, \aap, 496, 513

\bibitem[Moutou et al.(2011)]{moutou11} Moutou, C., Mayor, 
M., Lo Curto, G., et al.\ 2011, \aap, 527, A63

\bibitem[Naef et al.(2001)]{naef01} Naef, D., Mayor, M., 
Pepe, F., et al.\ 2001, \aap, 375, 205

\bibitem[Naef et al.(2001b)]{naef01b} Naef, D., et al.\ 2001, 
\aap, 375, L27

\bibitem[Naef et al.(2004)]{naef04} Naef, D., Mayor, M., 
Beuzit, J.~L., et al.\ 2004, \aap, 414, 351

\bibitem[Naef et al.(2007)]{naef07} Naef, D., Mayor, M., 
Benz, W., et al.\ 2007, \aap, 470, 721

\bibitem[O'Toole et al.(2007)]{otoole07} O'Toole, S.~J., Butler, R.~P., 
Tinney, C.~G., et al.\ 2007, \apj, 660, 1636

\bibitem[O'Toole et al.(2009a)]{otoole09a} O'Toole, S.~J., Tinney, 
C.~G., Jones, H.~R.~A., et al.\ 2009a, \mnras, 392, 641

\bibitem[O'Toole et al.(2009b)]{16417paper} O'Toole, S., Tinney, 
C.~G., Butler, R.~P., et al.\ 2009b, \apj, 697, 1263 

\bibitem[O'Toole et al.(2009c)]{monster} O'Toole, S.~J., Jones, 
H.~R.~A., Tinney, C.~G., et al.\ 2009c, \apj, 701, 1732

\bibitem[Peek et al.(2009)]{peek09} Peek, K.~M.~G., 
Johnson, J.~A., Fischer, D.~A., et al.\ 2009, \pasp, 121, 613

\bibitem[Pepe et al.(2011)]{pepe11} Pepe, F., Lovis, C., S{\'e}gransan, 
D., et al.\ 2011, \aap, 534, A58

\bibitem[Pepe et al.(2002)]{pepe02} Pepe, F., Mayor, M., 
Galland, F., et al.\ 2002, \aap, 388, 632

\bibitem[Perrier et al.(2003)]{perrier03} Perrier, C., Sivan, 
J.-P., Naef, D., Beuzit, J.~L., Mayor, M., Queloz, D., \& Udry, S.\ 2003, 
\aap, 410, 1039 

\bibitem[Robertson et al.(2012b)]{204313} Robertson, P., 
Horner, J., Wittenmyer, R.~A., et al.\ 2012b, \apj, 754, 50 

\bibitem[Robertson et al.(2012a)]{155358} Robertson, P., Endl, 
M., Cochran, W.~D., et al.\ 2012a, \apj, 749, 3

\bibitem[Rodigas \& Hinz(2009)]{rh09} Rodigas, T.~J., \& 
Hinz, P.~M.\ 2009, \apj, 702, 716

\bibitem[Santos et al.(2001)]{santos01} Santos, N.~C., Mayor, 
M., Naef, D., et al.\ 2001, \aap, 379, 999

\bibitem[Santos et al.(2008)]{santos08} Santos, N.~C., Udry, S., Bouchy, 
F., et al.\ 2008, \aap, 487, 369

\bibitem[Santos et al.(2010)]{santos10} Santos, N.~C., Mayor, M., Benz, 
W., et al.\ 2010, \aap, 512, A47

\bibitem[Santos et al.(2011)]{santos11} Santos, N.~C., Mayor, 
M., Bonfils, X., et al.\ 2011, \aap, 526, A112

\bibitem[Schwarz et al.(2009)]{schwarz09} Schwarz, R., S{\"u}li, {\'A}., 
Dvorak, R., \& Pilat-Lohinger, E.\ 2009, Celestial Mechanics and 
Dynamical Astronomy, 104, 69

\bibitem[S{\'e}gransan et al.(2010)]{seg10} S{\'e}gransan, 
D., Udry, S., Mayor, M., et al.\ 2010, \aap, 511, A45

\bibitem[Shen \& Turner(2008)]{shen08} Shen, Y., \& Turner, 
E.~L.\ 2008, \apj, 685, 553

\bibitem[Sozzetti et al.(2006)]{soz06} Sozzetti, A., Udry, 
S., Zucker, S., et al.\ 2006, \aap, 449, 417

\bibitem[Tamuz et al.(2008)]{tamuz08} Tamuz, O., 
S{\'e}gransan, D., Udry, S., et al.\ 2008, \aap, 480, L33

\bibitem[Tinney et al.(2003)]{tinney03} Tinney, C.~G., Butler, R.~P., 
Marcy, G.~W., et al.\ 2003, \apj, 587, 423

\bibitem[Tinney et al.(2005)]{tinney05} Tinney, C.~G., Butler, R.~P., 
Marcy, G.~W., et al.\ 2005, \apj, 623, 1171

\bibitem[Tinney et al.(2011a)]{tinney11a} Tinney, C.~G., Butler, R.~P., 
Jones, H.~R.~A., et al.\ 2011a, \apj, 727, 103

\bibitem[Tinney et al.(2011b)]{tinney11} Tinney, C.~G., Wittenmyer, 
R.~A., Butler, R.~P., Jones, H.~R.~A., O'Toole, S.~J., Bailey, J.~A., 
Carter, B.~D., \& Horner, J.\ 2011b, ApJ, 732, 31

\bibitem[Udry et al.(2002)]{udry02} Udry, S., Mayor, M., 
Naef, D., et al.\ 2002, \aap, 390, 267

\bibitem[Vogt et al.(2002)]{vogt02} Vogt, S.~S., Butler, R.~P., Marcy, 
G.~W., et al.\ 2002, \apj, 568, 352

\bibitem[Vogt et al.(2010)]{61vir} Vogt, S.~S., et al.\ 2010, 
\apj, 708, 1366 

\bibitem[Wittenmyer et al.(2007)]{witt07} Wittenmyer, R.~A., Endl, M., 
Cochran, W.~D., \& Levison, H.~F.\ 2007, \aj, 134, 1276

\bibitem[Wittenmyer et al.(2009)]{witt09} Wittenmyer, R.~A., Endl, M., 
Cochran, W.~D., Levison, H.~F., \& Henry, G.~W.\ 2009, \apjs, 182, 97

\bibitem[Wittenmyer et al.(2010)]{foreverpaper} Wittenmyer, R.~A., 
O'Toole, S.~J., Jones, H.~R.~A., Tinney, C.~G., Butler, R.~P., Carter, 
B.~D., \& Bailey, J.\ 2010, \apj, 722, 1854

\bibitem[Wittenmyer et al.(2011a)]{jupiters} Wittenmyer, R.~A., Tinney, 
C.~G., O'Toole, S.~J., Jones, H.~R.~A., Butler, R.~P., Carter, B.~D., 
\& Bailey, J.\ 2011a, \apj, 727, 102

\bibitem[Wittenmyer et al.(2011b)]{etaearth} Wittenmyer, R.~A., Tinney, 
C.~G., Butler, R.~P., et al.\ 2011b, \apj, 738, 81

\bibitem[Wittenmyer et al.(2011c)]{47205paper} Wittenmyer, R.~A., Endl, 
M., Wang, L., et al.\ 2011c, \apj, 743, 184

\bibitem[Wittenmyer et al.(2012a)]{HUAqr} Wittenmyer, 
R.~A., Horner, J., Marshall, J.~P., Butters, O.~W., \& Tinney, C.~G.\ 
2012a, \mnras, 419, 3258

\bibitem[Wittenmyer et al.(2012b)]{142paper} Wittenmyer, R.~A., 
Horner, J., Tuomi, M., et al.\ 2012b, \apj, 753, 169 

\bibitem[Wittenmyer et al.(2012c)]{24Sex} Wittenmyer, R.~A., 
Horner, J., \& Tinney, C.~G.\ 2012c, \apj, 761, 165

\bibitem[Wittenmyer et al.(2013)]{witt13} Wittenmyer, R.~A., Tinney, 
C.~G., Horner, J., et al.\ 2013, \pasp, 125, 351

\bibitem[Wright et al.(2009)]{wright09} Wright, J.~T., Upadhyay, S., 
Marcy, G.~W., et al.\ 2009, \apj, 693, 1084

\bibitem[Wright et al.(2011)]{wright11} Wright, J.~T., et al.\ 2011, 
\pasp, 123, 412

\bibitem[Zhou et al.(2007)]{zhou07} Zhou, J.-L., Lin, 
D.~N.~C., \& Sun, Y.-S.\ 2007, \apj, 666, 423 


\end{thebibliography}
\end{document}